\documentclass[amsmath,superscriptaddress,twocolumn]{revtex4-2}
\usepackage{enumitem} 
\usepackage{siunitx}
\usepackage{graphicx}
\usepackage{xcolor}
\usepackage{lipsum}
\usepackage{siunitx}
\usepackage[colorlinks=true,citecolor=blue,urlcolor=blue,linkcolor=blue]{hyperref}
\begin{document}
\setcitestyle{super}

\title{Phase Behavior of Unilamellar Hybrid Lipid-Diblock Copolymer Membranes}
\author{James F. Tallman}
\author{Junyu Wu}
 \author{Antonia Statt}
  \affiliation{Department of Materials Science and Engineering, The Grainger College of Engineering, University of Illinois Urbana-Champaign, Urbana, Illinois 61801, USA}
\begin{abstract}
    Hybrid lipid block copolymer membranes are promising for many applications in drug delivery, single molecule detection, in-membrane protein folding, and synthetic cells. 
    However, rational design is difficult due to the many design parameters which determine the nano- and micron-scale morphology and properties. 
    In this work, we propose a physically-informed framework which incorporates chemical immiscibility, hydrophobic thickness mismatch and geometric constraints to predict the morphology of hybrid membranes. 
    For this purpose, we extend existing theory for amphiphilic monolayers to model the thickness of diblock copolymer bilayers, demonstrating that both the hydrophobic and  hydrophilic block lengths determine the thickness. 
     We identify and rationalize the four primary membrane morphologies observed: mixed, laterally phase-separated, unzipped (thick–thin coexistence), and polymer-rich. 
     Specifically, chemical immiscibility differentiates mixed membranes from laterally phase separated membranes, and hydrophobic mismatch drives transitions to unzipped or polymer-rich morphologies. Areal density, finally, determines the crossover between unzipped and polymer-rich states. 
     We validate our theoretical predictions using coarse-grained molecular dynamics
    across a broad parameter space, including multiple lipid species (DOPC, DPPC), polymer species (1,4 PBD-$b$-PEO, 1,2 PBD-$b$-PEO, PE-$b$-PEO), block lengths, temperatures, and compositions.
     The resulting phase maps unify previously reported experimental and simulation observations and enable a generic and mechanistic understanding for the effect of system parameters on the nanoscale morphology.
\end{abstract}
\maketitle

\section{Introduction}
Hybrid lipid--block copolymer membranes combine the bio-compatible and functional properties of lipid membranes \cite{barenholz1993quality, bangham1965action} with the mechanical robustness of polymer membranes. \cite{discher1999polymersomes,Discher2002,rideau2018liposomes,le2011recent}
These hybrid lipid-polymer membranes show highly tunable, synergistic behavior which can exceed the properties of their components \cite{Go2021, Schulz2015, DeLeo2021}. Broadly, these systems are being pursued for applications in drug delivery \cite{Tuteja2018analyst,matsumoto2015therapeutic,martinez2025composition}, enhanced bio-molecular sensing \cite{vreeker2025nanopore}, the creation of synthetic cells \cite{maruvsivc2020constructing,Nishimura2020,otrin2021route, de2025distinct}, and biophysical understanding \cite{Steinkuhler2022}.

Many structures and morphologies have been observed in hybrid membranes. Diffraction limited characterization techniques (e.g. fluorescence microscopy) report `homogeneous mixing' and `phase separation' in hybrid membranes \cite{Go2021}. 
Higher-resolution techniques \textemdash neutron scattering \cite{fauquignon2021large,perera2022nanoscale,dao2017mixing},  cryo-electron microscopy \cite{Seneviratne2023,maruvsivc2020constructing,Tallman2024,fauquignon2021large,perera2022nanoscale,presutti2025balancing,dao2017mixing,Kambar2023softmat}, high resolution AFM \cite{di2022tailoring}, and molecular dynamics simulation \cite{Hu2019,schlitter2025nanoscale,vreeker2025nanopore,Steinkuhler2022,Muller2023} \textemdash reveal at least four distinct morphologies: mixing, unzipping, lateral phase separation, and polymer-rich membranes. 
The origin of these different morphologies is chemical immiscibility and hydrophobic mismatch (difference in thickness between the lipid and polymer hydrophobic cores) \cite{Go2021} which result in lateral phase separation and vertical phase separation (polymer core with lipid coating).

These membrane structures define the membrane function. For instance, finite-sized lateral domains could provide lipid rafts suitable for protein incorporation \cite{vreeker2025nanopore} while polymer-rich and unzipped domains can maximize lipid-polymer interfacial area, stabilizing low solubility drug molecules \cite{Tuteja2018analyst,Kambar2023softmat}. Vesicles coated with PEG are bio-compatible because they repel proteins from fouling to the surface\cite{suk2016pegylation}; hybrid lipid-polymer vesicles innately have a PEG coating only if the distribution of polymers is homogeneous (polymer-rich, mixed morphologies). Understanding the thermodynamics underlying these structures is therefore necessary for rational design.

Prior work has begun connecting molecular properties to emergent morphologies. In a review of experimental literature, Leal and Go identified hydrophobic thickness mismatch and Hansen solubility parameters as key indicators for macroscopic phase separation or mixing\cite{Go2021}. Hu and coworkers utilized dissipative particle dynamics (DPD) simulations to show that hydrophobic thickness mismatch and weak chemical incompatibility from the hydrophobic layer drive the resulting membrane morphology between mixing, lateral phase separation, and lipid rich phases. \cite{Hu2019} Schlitter and coworkers utilize Martini simulations to characterize the morphology resulting from using different lengths and concentrations of 1,2 PBD-$b$-PEO \cite{schlitter2025nanoscale}, showing mixing, unzipped, and polymer-rich membranes. Notably unzipped membranes are discussed as `nanoscale heterogeneity,' which appear in systems with low concentration of dilute polymers \cite{schlitter2025nanoscale}. Nevertheless, the physical origin of certain morphologies, their transitions, and the relation to the suite of properties intrinsic to lipids and block copolymers is still developing, which hinders the ability to rationally design hybrid membranes for specific applications.

In this work, we propose a generic framework which predicts the four morphologies based on hydrophobic mismatch and chemical immiscibility. Coarse-grained molecular dynamics simulations are used to validate this framework across an expansive parameter space: polymer block lengths (5-40 monomers), lipid types (DOPC, DPPC ($L_\alpha$), DPPC ($L_\beta$), polymer types (1,4 PBD-$b$-PEO, 1,2 PBD-$b$-PEO, PE-$b$-PEO), temperature (2.25 $k_BT$ to 3.75 $k_BT$) and concentration (10 to 95 mol \%) of polymers. 
Additionally, we extend amphiphilic monolayer theory \cite{wang1991curvature} to predict polymer membrane thickness as a function of both blocks. This enables understanding how hydrophilic block length modulates hydrophobic membrane thickness and the resulting morphologies.
This work demonstrates that competing interactions at the $nm$ length scale --- entropy, chemical immiscibility and hydrophobic mismatch --- can be tuned to create well-defined and structured materials.
The understanding of these interactions provides a rational design framework useful for obtaining a morphology of interest given a selected lipid and polymer. 

\section{Results and Discussion}

\subsection{Hydrophobic and Hydrophilic Blocks Define Pure Polymer Bilayer Properties}
\label{sec:single}

\begin{figure*}
    \centering
    \includegraphics[width=.95\linewidth]{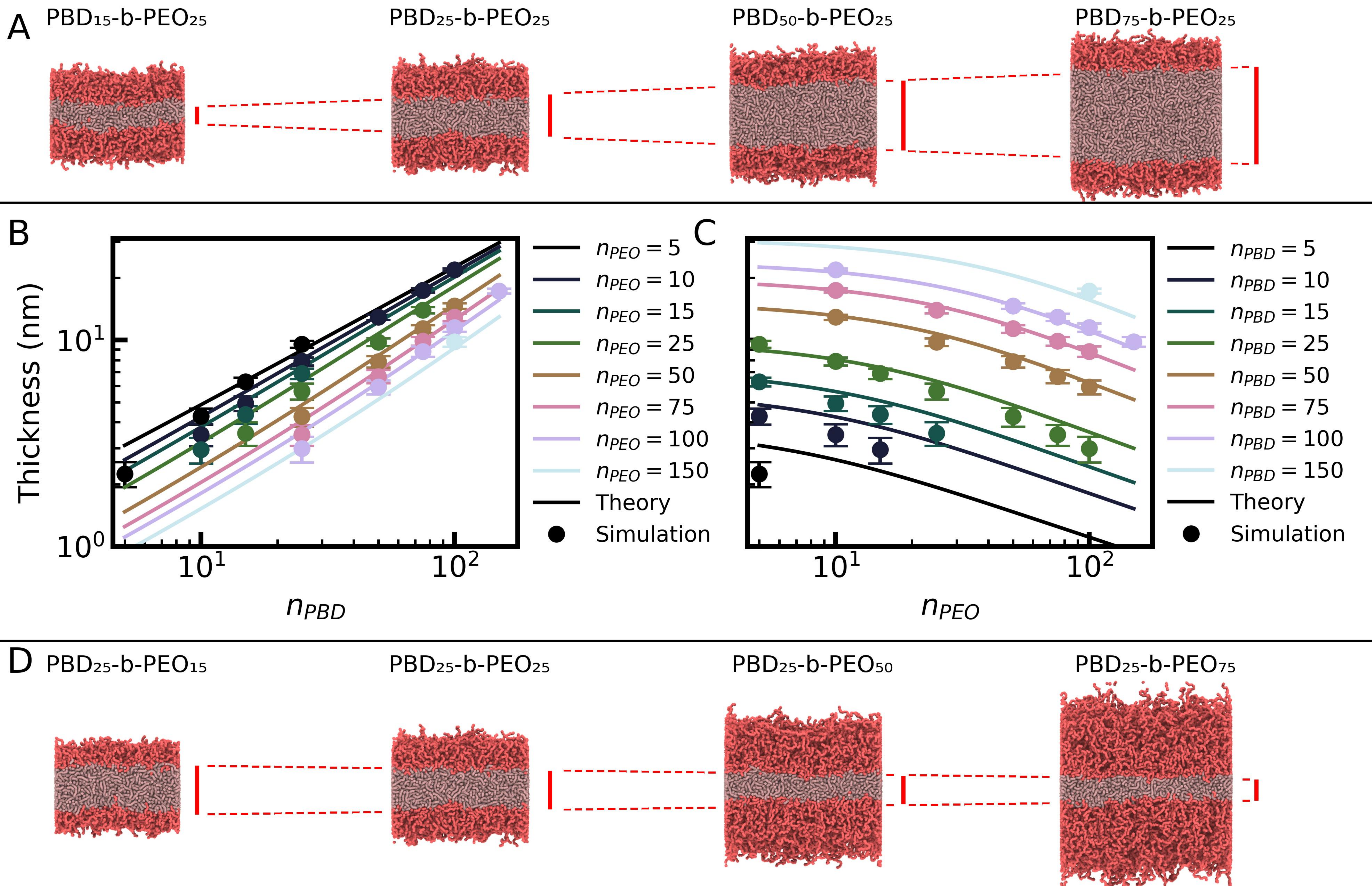}
    \caption{(A) Snapshots of membranes with constant PEO block length (25) and increasing PBD block length from left to right. Solid red vertical lines beside bilayers indicate the hydrophobic thickness of membranes, which increase with increasing PBD block length. (B) Thickness of hydrophobic core of amphiphilic polymer bilayers predicted from theoretical model (lines) and simulated data (points). (C) Thickness of membranes as PBD is increased for constant PEO lengths. (Right) Thickness of membranes as the PEO block length is increased for constant PBD block lengths. (D) Snapshots of membranes with constant PBD block length (25) and increasing PEO length from left to right. Solid red vertical lines beside bilayers indicate the hydrophobic thickness of membranes, which decreases with increasing PEO length.}
    \label{fig:pure-polym-thickness}
\end{figure*}

The thickness and area per polymer chain of diblock copolymer membranes depends both on the length of the hydrophilic and hydrophobic chain lengths (Figure \ref{fig:pure-polym-thickness}).
However, the thickness of amphiphilic polymer bilayers is frequently written as a power law scaling only with respect to the molecular weight of the hydrophobic block \cite{bermudez2002molecular,srinivas2004self,battaglia2005bilayers,grillo2017diblock,kowalik2017chemically,barden2018parameterization,ortiz2005dissipative,le2011recent}. This approach does not model the effect of the hydrophilic polymer brush (although it is implicitly handled by maintaining a constant ratio of PBD to PEO).
On the other hand, it has been suggested that a range of scaling exponents could be expected based on the importance of the hydrophilic polymer brush\cite{ortiz2005dissipative}. The thickness of the hydrophilic corona has been shown to scale linearly with the PEO degree of polymerization, suggesting that it is a polymer brush \cite{smart2009polymersomes}.
Wang and Safran \cite{wang1991curvature} put forth a theory for amphiphilic monolayers with a hydrophilic chain in good solvent and hydrophobic chain in melt-like conditions, of which the zero curvature limit is relevant to bilayers.

We extend this model\cite{wang1991curvature} to polymer bilayers with several assumptions: to the first order, polymer bilayers consist of uncoupled polymer monolayers, implying the end-to-end distance of the polymer, $R$, is proportional to half the thickness of the bilayer, $d/2$; a mean field description of the polymer brush is sufficient to model the polymer brush \cite{milner1988theory}; and chain confinement is relevant. The free energy per chain in our formulation can be expressed as a sum of the interfacial tension, the energy associated with stretching and confining the hydrophobic polymer chain, and the energy associated with the PEO block being in a polymer brush, or, 
\begin{equation}
\begin{aligned}
F \approx\;& \frac{\gamma}{\sigma}  + kT \Bigg[
    \frac{3}{2}\,\frac{n_b}{C_{\infty,b}}\, a_b^4\, \sigma^2
    + \frac{C_{\infty,b}}{2 n_b\, a_b^4\, \sigma^2}
\\
&
    + \frac{9}{10}\!\left(\frac{\pi^2}{12}\right)^{1/3}
      \left(
        \frac{
            \, (C_{\infty,a}^{xy} c_a)^2\, \sigma
        }{
            C_{\infty,a} 
        }
      \right)^{2/3}
      \frac{n_a}{C_{\infty,a}}
\Bigg] \quad .
\end{aligned}
\end{equation}

In these equations, the subscript $a$ denotes the hydrophilic chain and $b$ denotes the hydrophobic chain. $N$ is the number of Kuhn monomers in the chain, $n$ is the number of monomers in a chain, $\sigma$ is the chains per monolayer per area, $a$ is the monomer size (the monomer volume $v$ is given by the length $a$ such that the volume $v = a^3$), $\gamma$ is the interfacial tension, and $C_\infty$ is the characteristic ratio. The excluded volume $v$ can be expressed as $\pi c_a^2b_a$ where $c_a$ is the radius of the excluded volume cylinder. $C^{xy}_{\infty,a} $ describes the anisotropic characteristic ratio of polymer chains (used to normalize the grafting density to statistical segments).

We extract the parameters for the free energy from: definitions of the Martini model ($kT$, $a_a$,$d_a$), fitting the polymer brush distributions to a parabolic distribution ($C^{xy}_{\infty,a}, c_a, C_{\infty,a}$, shown in Figure S2), and measuring the density and chain stiffness of polymer melt ($a_b, C_{b,\infty}$). The interfacial tension is the only free parameter that is directly fitted to the simulated data ($\gamma = 11.73$ \unit{\kJ \per \mol \per \nm \squared}). All parameters are shown in Table S2. A more granular discussion on the extraction of parameters can be found in the supplementary information. The free energy is minimized with respect to the grafting density $\sigma$ to obtain the equilibrium grafting density. Finally, the thickness is calculated with the relationship $d = 2\sigma n_b a_b^3$.

As shown in Figure \ref{fig:pure-polym-thickness}, there is excellent agreement between the theoretical model and the simulated results. A parity plot of the predicted and simulated thickness confirms this (Figure S3).  However, notably, the model fails to model membranes with very small PEO or PBD block lengths. This is expected, as the effects of chains with fewer than several Kuhn segments are not modeled by the free energy expression. This model is also not expected to hold in highly curved bilayers (Wang and Safran expanded the free energy for varying radii of curvature \cite{wang1991curvature}). 

Some interesting behaviors arise from this model:
In the limit of a short PEO brush, only chain stretching and interfacial energy enter the free energy expression, resulting in the strong segregation limit  ($d\propto n_{PBD}^{2/3}$).
When $n_{PEO}> 50$, at constant PEO, $d \propto n^1_{PBD}$.
When $n_{PBD} > 50$ and $n_{PEO}>100$, increasing PEO yields a power law-like behavior with a relationship of $d\propto n_{PEO}^{-0.34}$.
Most notably, when PEO and PBD are increased simultaneously, there is no obvious power law scaling (as shown in Figure S5). 
The purely elastic stretching and bending moduli can be approximated from the free energy expression (Figure S6). Membranes are predicted to be stiffest when either $n_{PEO}\gg n_{PBD}$ or $n_{PBD}\gg n_{PEO}$. When $n_{PBD} \sim n_{PEO}$, the modulus is predicted to be relatively independent of chain length, as is experimentally found \cite{bermudez2002molecular}, and slightly higher than the mean field prediction of $4\gamma \sim 80$ \unit{\mN\per \m}.

Here, we highlight that the thickness of a polymerosome (consisting of a bilayer of hydrophobic and hydrophilic polymers) is dependent on both the length of the hydrophobic and the hydrophilic block. The free energy from which the membrane thickness is derived is adapted from the model for amphiphilic monolayers \cite{wang1991curvature}, adding an energy for chain confinement and treating the bilayer as two uncoupled monolayers. The predicted thicknesses, with only one explicit fit parameter, agree strongly with the simulated results over a large range of PBD length, PEO length, and membrane thicknesses. This demonstrates that the theory -- as alluded to as a possibility by Ortiz and coworkers \cite{ortiz2005dissipative} -- is a reasonable model to apply to the system of diblock copolymers bilayers. 

\subsection{Morphologies of a Hybrid System}
\label{sec:morph}
There are four morphologies that appear in symmetric unilamellar hybrid bilayer membranes which we have observed in this work and prior literature. 
Each of these morphologies has unique properties, applications, and appears under different conditions.
Asymmetric \cite{huang2026mechanical}, highly curved \cite{Tallman2024}, and non-lamellar \cite{kang2023cooperative} membranes can also be synthesized experimentally, but will not be discussed in this work. 
The morphologies are as follows.
\begin{description}[style=sameline, leftmargin=0.25cm, labelwidth=0.25cm]
\item[Mixing] Polymers (or lipids) interdigitate randomly in a matrix of the other type. This appears when there is minimal driving force for phase separation, as determined by the hydrophobicity of the molecules, and when the molecules are relatively short. Mixed membranes can be useful because they induce packing defects in otherwise lipid-like membranes \cite{Steinkuhler2022}.
    
\item[Lateral Phase Separation (LPS)] Polymer-rich and lipid-rich domains form which have similar thickness, driven by the hydrophobicity difference. These domains can continue to coarsen by fusing or by ripening. These membranes can have mechanical stability of polymer-rich membranes while maintaining the bio-compatibility of lipid membranes \cite{vreeker2025nanopore}.
    
\item[Unzipped/Thick and Thin Phases] polymer-rich globules can form inside of lipid monolayers \cite{Tallman2024,schlitter2025nanoscale}. Unzipped domains appear in systems with constrained number of molecules (vesicles, finite sized simulations), having unique behavior (increased continuity of monolayers, larger protrusions) compared to the thick-and-thin phases. Thick-and-thin phases also appear in experimentally synthesized systems \cite{Tallman2024}, and have similar features: continuous lipid bilayer in coexistence with a polymer-rich bilayer where lipids coat the surface. 
Unzipped domains and the coexistence of thick and thin phases are in the same category because of arguments put forth in Section \ref{sec:unzipping}. In summary, for a membrane to sustain a large hydrophobic thickness mismatch, the interface is several $nm$ large. When a small domain, which would become coexisting thick and thin phases in the thermodynamic limit, has a similar size to that of the interfacial width, it appears unzipped.
    
\item[Polymer-rich] A polymer-rich vesicle has the character of a polymerosome with lipids acting as surfactants, attracted to the interface of the hydrophobic polymer-molecules and the water. These membranes are frequently observed \cite{Tallman2024,Seneviratne2023} under cryo-electron microscopy, are homogeneous, and have the mechanical properties similar to pure polymer bilayers.
\end{description}

The coexistence of thick and thin phases is conceptually similar to lateral phase separation, both are forms of phase separation; however, we draw the distinction because lateral phase separation has no thickness mismatch, displays a melting transition to a mixed morphology at constant polymer lengths, and no lipids float atop laterally phase separated domains. Also, the interfaces between laterally phase separated domains are molecularly thin, whereas coexisting thick and thin domains have large interfaces due to unzipping. 

For each transition between two morphologies, we propose a predictive physical principle from system parameters which explains the transition. We then show that these principles agree with simulation by construction of a thoroughly explored phase-map (Figure \ref{fig:classification}).

\subsubsection{Mixing and Lateral Phase Separation Follows Flory Huggins Spinodal}
The transition from mixing to lateral phase separation can be described with a Flory-Huggins type model\cite{rubinstein2023polymer}.  This expresses the free energy of mixing as
\begin{equation}
    F_{mix} = \frac{\phi}{N_{lipid}} \ln\phi + \frac{(1-\phi)}{N_{PBD}}\ln(1-\phi) + \chi \phi(1-\phi)\quad.
\end{equation}
The spinodal can be constructed from the roots of the second derivative of the free energy.
To account for the connectivity of lipid and polymers and treat this as a 2D phase separation, we map the lipid and polymer with the relationship $N = n/c$. We select $c = 4$ because a lipid with two tails (consisting of 4 beads each) can be represented as two connected beads on a top-down 2D lattice, or, $N_{lipid} = 2$. A PBD chain with 20 monomers is then treated with $N_{PBD} = 5$, maintaining the same relationship. The enthalpic term is scaled by this coarse-gaining factor to account for the interactions between all 4 beads.    

The Flory-Huggins $\chi$ parameter is
defined by the coordination number, thermal energy, and pairwise interaction energies, which can be calculated from the model parameters (Table S3).
As shown in Figure \ref{fig:classification}, the spinodal describes the data from the model system with reasonable accuracy, predicting mixing on the outside of the spinodal and lateral phase separation inside the spinodal. One point is sampled to the right of the spinodal which does not display mixing. This can be attributed to second-order effects which extend beyond chemical immiscibility. For instance, there is non-ideal mixing due to the nematic ordering of lipids in the membrane which is interrupted by the inclusion of polymers. These effects are not included in the FH free energy expression. 

Chemical immiscibility can be removed from the simulation by replacing the polybutadiene molecule (a $C4$ type bead in the Martini parameterization) with a polyethylene-like molecule (which uses $C1$ type beads, the same as unsaturated lipid tails).
The resulting simulated morphologies are shown in Figure S15. 
Upon removing the chemical immiscibility, lateral phase separation completely disappears. This shows that lateral phase separation occurs only due to a chemical incompatibility between (relatively short) block copolymers and lipids.

\subsubsection{Hydrophobic Mismatch Differentiates Mixing and LPS from Unzipped and Polymer-rich Morphologies}
Hydrophobic mismatch is one of the two parameters identified in the review by Go and Leal \cite{Go2021} which determine the mixing behavior of hybrid lipid--polymer membranes. Hydrophobic mismatch is defined to be the difference in the equilibrium polymer membrane thickness and the equilibrium lipid membrane thickness. In section \ref{sec:single}, we developed a model for the thickness of pure polymer membranes. When the pure polymer membrane becomes thicker than the pure lipid membrane, the membranes shift from lateral phase separation or mixing to unzipping or polymer-rich morphologies. This threshold is indicated by the horizontally dashed lines in Figure \ref{fig:classification}. Three lines are shown for differing PEO lengths, indicating that the hydrophilic chain length can impact the membrane morphology.

This unzipping and polymer-rich behavior originates because of the shape and stiffness of the constituent molecules. As shown in Figure S15, systems with no hydrophobicity difference between lipid tails and polymers still display unzipping and polymer-rich morphologies. Purely hydrophobic molecules such as monomers in the parachute morphology \cite{jung1997new,Jung2000vesicle,Hubert2000vesicle}, fats in budding lipid droplets\cite{walther2017lipid}, nitrogen accumulation, hydrophobic nanoparticles \cite{VonWhite2012structural,Marzouq2024} and hydrophobic polymers \cite{bochicchio2017} also display unzipped domains. Lipid molecules have wide head and tails, whereas polymers have narrow-but-long blocks. It is favorable for the polymer to occupy the interior of the membrane to maintain nematic order of lipids and the lipid-water interface.

\subsubsection{Areal Density Determines the Crossover Between Unzipped and Polymer-Rich Morphologies}
\begin{figure*}
    \centering
    \includegraphics[trim={0cm 0cm 0cm 0cm},clip,width=1\linewidth]{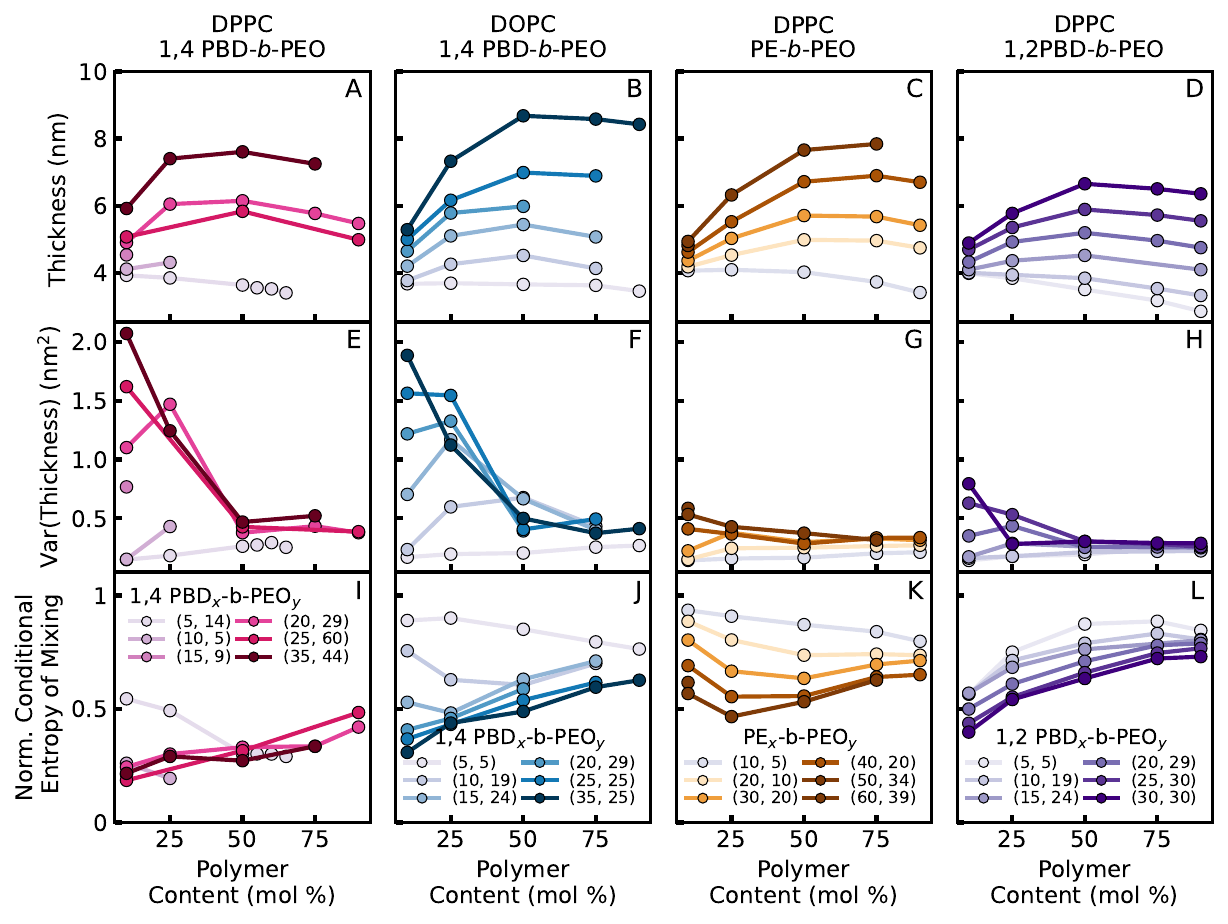}
    \caption{Properties of hybrid membranes for different combinations of lipids and polymers, as function of polymer content ($x$ axis) and hydrophobic chain length (color). (A-D) Thickness of the hydrophobic core of the membrane. (E-H) Spatial variance of thickness of hydrophobic core of the membrane. (I-L) Normalized conditional entropy of mixing between hydrophobic lipid tail beads and hydrophobic polymer beads. (A,E,I) DPPC and 1,4 PBD-$b$-PEO. (B,F,J) DOPC and 1,4 PBD-$b$-PEO, (C,G,K) DPPC and PE-$b$-PEO. (D,H,L) DPPC and 1,2 PBD-$b$-PEO.}
    \label{fig:12pbd}
\end{figure*}

The transition between unzipping and polymer-rich morphologies can also be predicted from the pure membrane properties. We assume lipid membranes are (relatively) incompressible and that polymer membranes have similar  thickness and area per chain between their pure states and when in hybrid membranes (Figure \ref{fig:12pbd}). As a result, the transition between unzipping and polymer-rich morphologies occurs when the area of the polymer phase exceeds that of the lipid phase, or,
\begin{equation}
    a_{polymer}\cdot\phi> a_{lipid}(1-\phi)\quad,
    \end{equation}
where $a_i$ is the area per monolayer per molecule and $\phi$ is the mole fraction of polymer. $a_{polymer}$ is equivalent to $1/\sigma$, where $\sigma$ is the chains per area discussed in Section \ref{sec:single}. These lines, which span vertically, are predicted from the modeled area per chain of the pure polymer membranes and drawn on Figure \ref{fig:classification} for different PEO lengths (indicated by different dashed lines). As with the hydrophobic mismatch, changing the length of the PEO changes exact value at which this transition is expected (as is also seen in simulation).
It is important to note that this geometric argument to separate polymer-rich vesicles and unzipped vesicles is only a guideline representing the maximum polymer content a bilayer can incorporate before being deemed polymer-rich. As discussed in Section \ref{sec:unzipping}, there is also a variable concentration of lipids which coat unzipped domains that depends on the size of the domain, the hydrophobicity of the lipid and the polymer, and the hydrophobic mismatch. Similarly, the lipids affect the thickness and the areal density of the polymers. Because of these two effects, this transition can move.

\subsection{Phase Map for a Model System}
\begin{figure*}
    \centering
    \includegraphics[width=.95\linewidth]{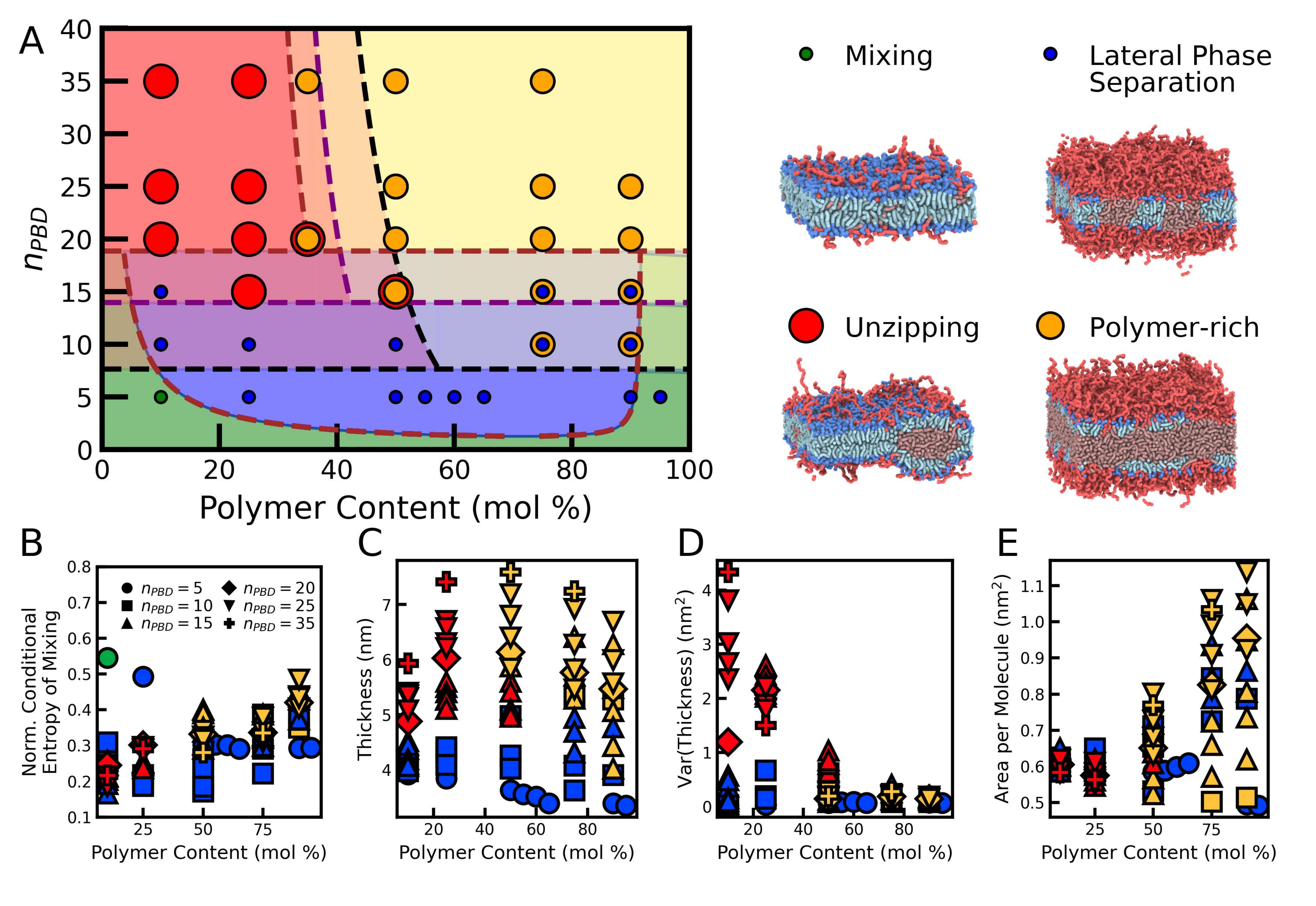}
    \caption{(A) Morphologies of hybrid DPPC (L$_\alpha$)--block co-polymer membranes as a function of polymer content and PBD block length. Shading of regions corresponds to predictions from theoretical expectations, where the boundaries between regions are a function of PEO block length ($n_{PEO}=5$: black dashed line, $n_{PEO}=25$: purple,$n_{PEO}=45$: red), mixed morphology are expected in green regions, lateral phase separation in blue regions, unzipping in red regions, and polymer-rich morphologies in yellow regions. Circle colors reflect morphologies observed from simulations at given state points. (right) Renderings of representative morphologies (lipid tails are colored light blue, lipid head groups: dark blue, PBD Monomers: light red, PEO monomers: dark red).
    (B-E) Properties of interest as a function of polymer content for all simulated DPPC and PBD-$b$-PEO membranes. Marker color indicates the final morphology. Marker shape represents the PBD block length. (B) Normalized conditional entropy of mixing. (C) Thickness of the hydrophobic core of the membrane. (D) Spatial variance of thickness over the membrane. (E) Area per molecule.}
    \label{fig:classification}
\end{figure*}

A model system with a fluid phase lipid (dipalmitoyl phosphatidylcholine, DPPC, at 298K, which is in the fluid phase in the Martini v3.0 model \cite{Souza2021}) and an amphiphilic block copolymer (1,4 polybutadiene-$b$-polyethylene oxide, PBD-$b$-PEO) initialized in a randomly mixed bilayer-like configuration is simulated at varying polymer content fractions, PBD lengths, and PEO lengths. The resulting morphologies are classified as being mixed, unzipped, laterally phase separated, or polymer-rich, as shown in Figure \ref{fig:classification}). Details of classification are presented in the methods. The phase map indicates the morphologies observed from simulations at a given concentration and PBD block length. The general behavior is as follows: with a short polymer at a dilute concentration, mixing is observed; short polymers at larger concentrations form laterally phase separated domains (where lipid regions have similar thickness to polymer regions); long polymers at dilute concentrations form `unzipped' domains where there is a continuous monolayer of lipids coating phase separated polymer globules; and long polymers at high concentrations form polymer-rich, thick bilayers.

Multiple morphologies can be observed at the same point because multiple PEO lengths are simulated for a given polymer content and PBD length. Lines are added to the phase map based on theoretical predictions and estimations. This includes the thickness of pure polymer membranes, thickness of pure lipid membranes, and a spinodal construction; these parameters are not tuned to the morphological behaviors themselves. The three sets of dotted lines indicate predicted boundaries for different PEO lengths.

Below the phase map, relevant properties of the hybrid membranes are shown with the simulated morphology indicated by the color of the marker. In (B), the normalized conditional entropy of mixing, which is a metric between zero (fully phase separated) and unity (perfectly molecularly mixed) that indicates how many contacts there are between lipid and polymer beads given the concentration of the system\cite{brandani2013quantifying}, is shown. 
The shortest polymers simulated at the most dilute concentrations have the highest conditional entropy of mixing. Mixing is suppressed by increasing either the concentration or chain length. At longer chain lengths, both unzipping and laterally phase separated domains have very similar conditional entropy of mixing. Both of these morphologies are phase separated: unzipping is in a vertical and horizontal direction, and lateral phase separation is in only horizontal direction, but the extent of demixing is similar. Finally, polymer-rich vesicles have an increased entropy of mixing compared to lateral phase separation because lipids can incorporate deeper into the membrane core.

In (C), the average thickness of the membranes is shown. Membranes with unzipped or polymer-rich domains are much thicker than membranes which laterally phase separate. In (D), the spatial variance of that thickness is monitored. The only morphology which has a large variance in thickness is unzipping, which consists of a coexistence of thick and thin domains. 
Finally, in (E), the area per molecule is shown, which is largely dependent on the polymer composition and does not correlate with a specific morphology.

\subsection{Nature of Mixing and Lateral Phase Separation Transition}

Above, we suggest that the relation between lateral phase separation and mixing is a mixing-demixing phase transition. Should this be the correct physical model for the behavior, there should exist a temperature where the system shifts from  macroscopic phase separation to mixing. As such, three randomly initialized simulations are cooled and heated repeatedly to observe the mixing behavior.
\begin{figure*}
    \centering
    \includegraphics[width=1\linewidth]{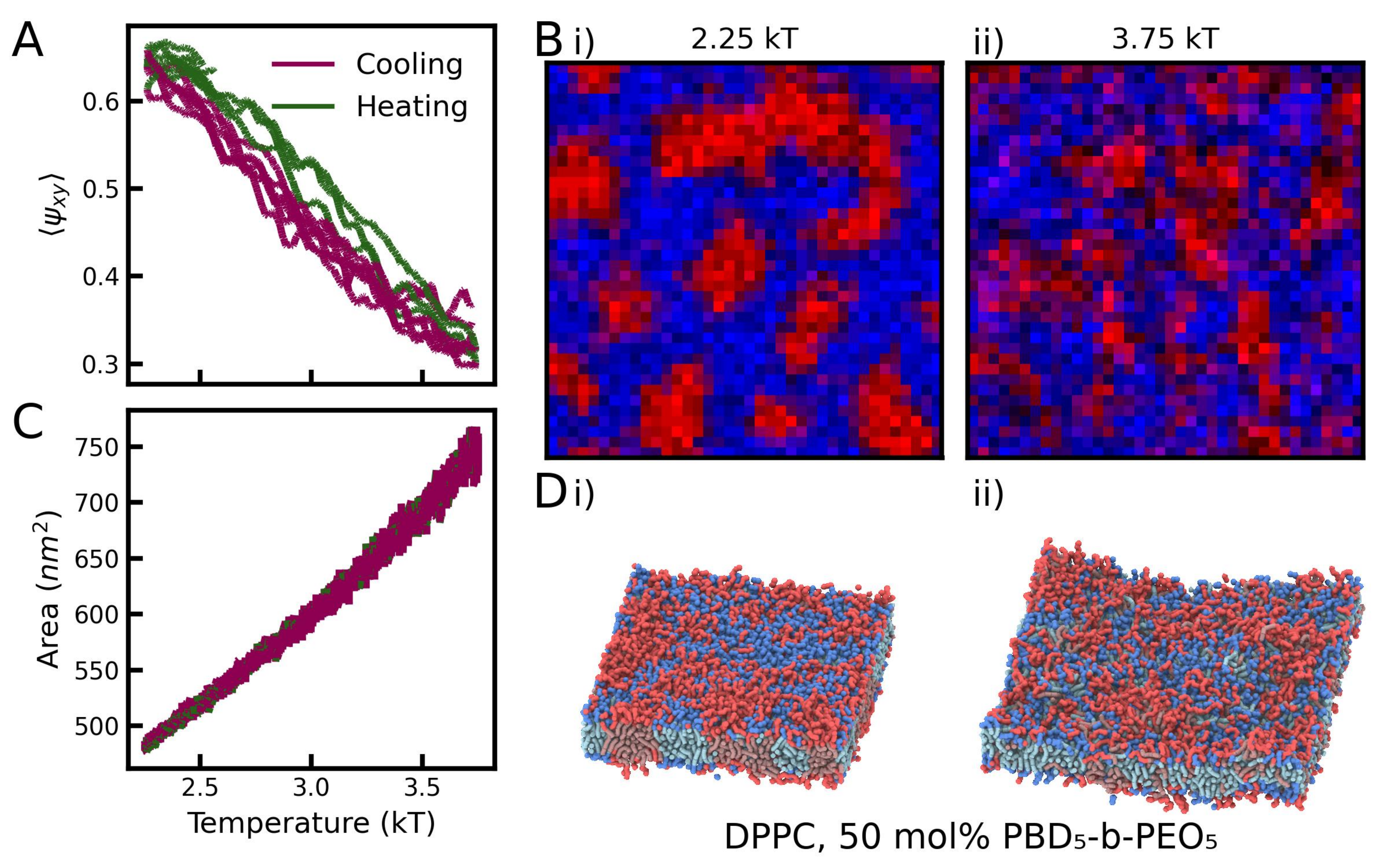}
    \caption{Investigation of temperature effect on phase separation of hybrid membranes. Three simulations performed on DPPC membranes with 50 mol \% PBD$_5$-b-PEO$_5$ were simulated for 18 $\mu$s with a heating/cooling rate of $0.75$ K/$\mu s$ . (A) Phase separation order parameter (0 if perfectly mixed, 1 if perfectly phase separated) plotted as a function of temperature. Order parameter is smoothed via convolution with window size of 20 frames. Green lines indicate data taken during heating and pink lines represent cooling. (B) Top Right: Histograms show positions of lipid tails (blue) and hydrophobic polymer blocks (red) at the low temperature (i) and high temperature (ii) extremes. (C) Area plotted as a function of temperature using the above color scheme.  These histograms were used to calculate the order parameter. (D) Snapshots of membranes at the low temperature (i) and high temperature (ii) limits.}
    \label{fig:temp_effect}
\end{figure*}

In Figure \ref{fig:temp_effect} (A), the mixing order parameter $\langle \psi_{xy}\rangle $\cite{jeong2024modeling}, which is unity if the system is perfectly phase separated and zero if the system is perfectly mixed, is shown. This order parameter varies between 0.65 at the lowest temperature (2.25 $k_BT$, 270K) and 0.3 at the highest temperature (3.75 $k_BT$, 450K). There is a slight hysteresis, the system tends to stay more phase separated when heating than cooling, consistent with a memory of the previous state. In (B), histograms of the lipid and polymer positions are shown (looking down on the membrane). The membranes at 2.25 $k_BT$ are clearly phase separated, also shown in (D), the order parameter does not go fully to unity because there are still many domains present in the material instead of full macroscopic phase separation. One reason that these domains do not coalesce during the simulated time is that the dynamics of the lipids are slowed dramatically at low temperatures. The phase separation can be further confirmed by analyzing the polymer-polymer pair correlation functions, which show a strong secondary valleys and peaks at low temperatures, consistent with domains, but no features beyond 2.5\unit{\nm} at high temperatures (Figure S13).

At high temperatures, the domains are less persistent and mixing is seen. Interestingly, the mixing is not perfect: polymers are still somewhat correlated with other polymers. These correlated regions are one mechanism to stabilize finite size clusters within a hybrid membrane, and could be engineered by selecting a lipid and polymer system with a transition temperature slightly below physiological temperatures.

Finally, in (C), the plot of area as a function of temperature is shown. The area of the membrane increases roughly linearly with temperature. There is no hysteresis upon heating or cooling, which indicates that there is no significant crystallization or gelation behavior. Additionally, three other simulations were performed at different concentrations and temperature ranges and show similar results (Figure S12). The simulation with only 10\% polymer does show hysteresis originating from a gel phase transition of the lipid which is suppressed by the inclusion of more polymer.

In summary, a system with short hydrophobic polymer and a fluid lipid that has reasonable hydrophobicity difference can undergo a mixing to demixing transition. Given the ability to modify the properties of the lipid phase by creating mixtures of lipids, it is possible to imagine a hybrid system which can be engineered to undergo a phase transition close to physiological transition, similar to liquid ordered - liquid-disordered phase transitions \cite{rayermann2017hallmarks}.

\subsection{Nature of Unzipping, Polymer-rich, and Coexistence of Thick and Thin Morphologies}
\label{sec:unzipping}
When the pure block copolymer membrane used has a preferred thickness greater than that of the lipid membrane, either an unzipped morphology or a polymer-rich morphology is observed in simulation. The above simulations were shown for relatively small systems ($\sim 24$ \unit{\nm}) initialized randomly.

Experimentally, these morphologies can also be observed. Both unzipped \cite{Tallman2024} and polymer-rich vesicles \cite{Seneviratne2023} have been characterized by scattering and cryo-EM imaging. Unzipped membranes have both lipid bilayers and regions where two continuous monolayers of lipids surrounding the polymer domains. Another morphology which has been observed is the coexistence of thick and thin domains in a single nanoparticle \cite{Tallman2024}. This appears different than unzipped domains due to the depletion of lipid on the surface of these polymer domains (observed by the decrease in contrast across the membrane). These two morphologies contain the same lipid and polymer species and differ solely in composition. This raises the question: Is unzipping the same as the coexistence of thick and thin phases? 

\begin{figure*}
    \centering
    \includegraphics[width=1\linewidth]{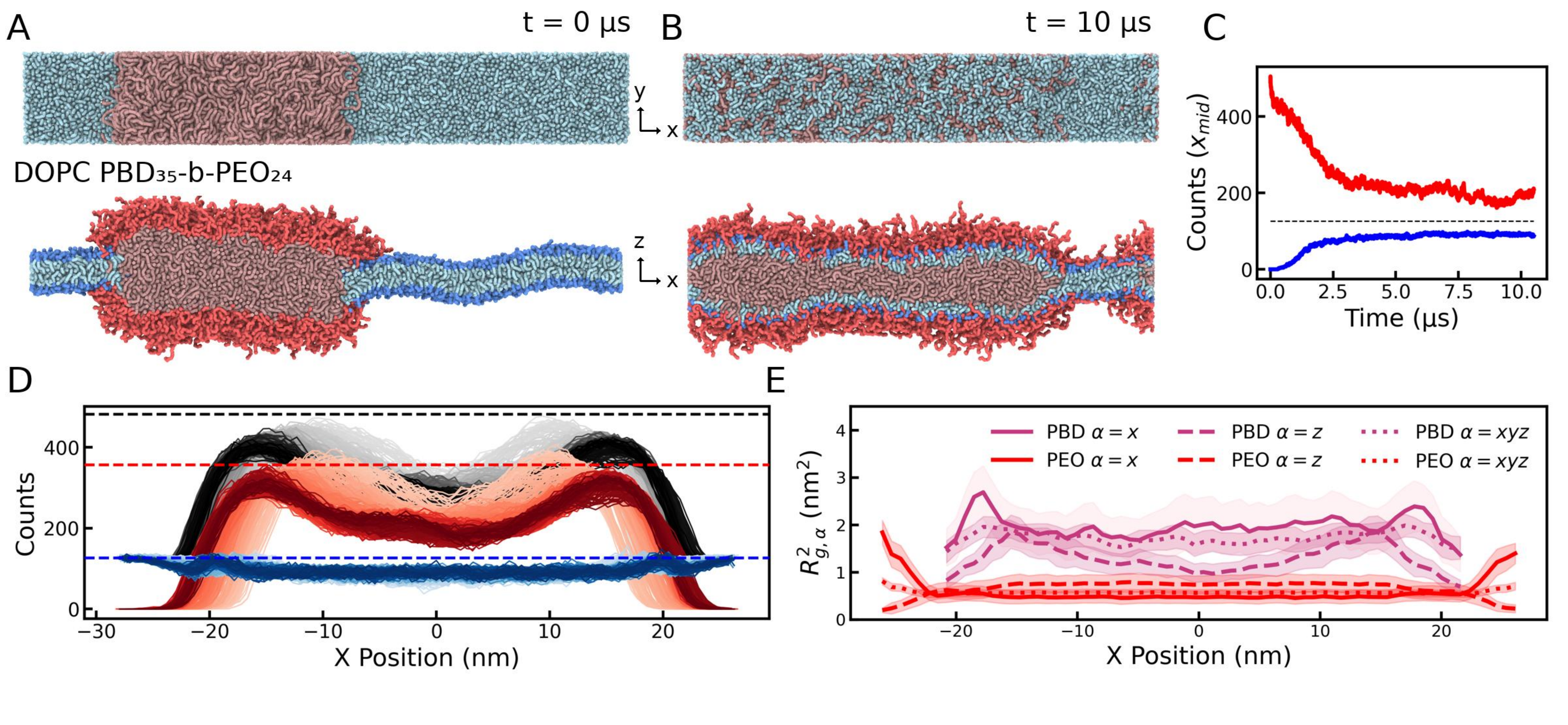}
    \caption{DOPC and PBD$_{35}$-b-PEO$_{24}$ hybrid membrane with 30 mol\% polymer simulated at 2.47 kT for 10 $\mu s$ under NP$_{xz}$LT ensemble. (A) Morphology of the system after initial equilibration, where the top snapshot shows a top down view of the membrane not showing the hydrophilic beads, and the second snapshot shows a side view of the membrane. (B) Morphology of the system after 10 $\mu$s of simulation from top down and side view. (C) Counts of lipids (blue) and polymers (red) measured in the center (x-coordinate) of the polymer domain as a function of simulation time. (D) Concentration of lipids (red), polymers (blue), and the sum of lipids and polymers (black) over the last 8 $\mu$s as a function of the x coordinate, histograms are generated over a window of five simulation frames. Dashed lines show the concentration of a pure lipid bilayer (blue), pure polymer membrane of the same PBD and PEO block lengths (red) and sum of pure lipid and pure polymer bilayers (black). (E) Components of the radius of gyration tensor of all polymers, in the x (solid line) and z (dashed line) dimension and total (dotted line), binned as a function of the x coordinate, for PEO (red) and PBD (purple) separately, averaged over the last 2 $\mu $s.}
    \label{fig:interface}
\end{figure*}

To investigate this, we initialize a large system (initially $\geq$ 30 \unit{\nm} lipid and polymer domains, shown in Figure \ref{fig:interface}) and allow it to equilibrate in an NP$_{xz}$LT ensemble, where the $y$ dimension of the box is set to a fixed length. The resulting morphology after equilibration (10 $\mu$s), shown in Figure \ref{fig:interface} (B), has similar features to the polymer-rich morphologies, unzipped morphologies, and coexistence of thick and thin phases. 

Lipids coat the polymer-water interface as seen in unzipped phases. This coating is about 80\% the concentration of a pure lipid monolayer; there is a concentration gradient of lipids across the interface. This gradient is shown in Figure \ref{fig:interface} (D) and the equilibration of the lipid concentration is shown in (C). We speculate that in a thicker membrane, a membrane with a more polar polymer, or a more ordered lipid, there would be a larger concentration gradient. One simulation (Figure S14) was performed with a more ordered lipid membrane, however the slower dynamics meant the simulation could not equilibrate on a feasible time scale.

Another unique behavior is that the concentration of both lipids and polymers in these hybrid membranes is largest at the unzipped interface. This concentration is directly proportional to an increase in membrane thickness. This increase in size at the interfaces is observed in both the density histograms (D) and the radius of gyration of polymer chains (E). One explanation for this effect is that of curvature. In (E), it is shown that the PEO chains at the edges of the membrane are enlarged in the $x$ direction and shrunk in the $z$ direction. This is due to the increased volume available for the polymer chains to explore around the curved interfaces. Since there is more volume for the polymer chain, the steric repulsion from the polymer brush has less of a thinning effect on the hydrophobic membrane. The corollary to this effect is that the PBD chains near the interface are also stretched in both directions compared to the polymer chains in the center of a polymer domain. The large concentration of lipids at the unzipped interface can be attributed to proximity to the semi-infinite source of lipids which diffuse onto the polymer domain.

The density of the polymer membrane with lipids coating the surface is lower than the density of the equivalent membrane simulated with no lipids. This implies that the membrane is thinner, likely due to the lipids decreasing the surface tension of the polymer-water interface. Nevertheless, at the unzipped domains, the total density (thickness) of lipids and polymers is greater than that of the pure polymer membrane. This explains why, at times, hybrid membranes can appear thicker than their pure polymer equivalents: lipids have a slight thinning effect on the polymer membrane, but also add height by themselves coating the surface of the membrane.

These effects are important for rationalizing several behaviors regarding the difference between unzipped and coexisting thick and thin domains. First, the interfaces of thick polymer-rich domains and thin lipid-rich domains are likely unzipped. We see this effect with both a disordered DOPC molecule and a more ordered (but not gel phase) DPPC molecule (Figure S14). Second, there is likely little physical difference between an unzipped domain and the coexistence of thick and thin phases. The interface between thick and thin domains must be unzipped to accommodate the substantial hydrophobic thickness mismatch. With sufficient polymer content, there is both an interfacial region and a bulk region. One reason that these appear so different in Cryo-EM imaging is the microfluidic self assembly; molecules form vesicles which then reorganize into polymer-rich and lipid rich domains, effectively creating a constant number of particles and fixed lipid mol fraction ensemble inside of each vesicle\cite{Tallman2024}. As a result, domains cannot coarsen indeterminately but rather are limited to some finite size. Thus, unzipped domains have a higher concentration of lipids because they are this interfacial region, whereas, coexistence of thick and thin phases can have fewer lipids because they are closer to the thermodynamic equilibrium of the two bulk phases.

\subsection{Effects of Polymer and Lipid Type on Phase Behavior}

The origin of the morphological behavior and the underlying properties are generic to the features of the lipid and the polymer; the explanation relies only on hydrophobic thickness mismatch and chemical immiscibility. The behavior of the system with respect to concentration of lipids and polymer lengths should be consistent, except for extreme cases. To confirm this behavior, and compare results with systems simulation in prior reports, simulations of increasing polymer lengths and increasing concentration (x-axis) were performed for the systems of DOPC + 1,4 polybutadiene-$b$-PEO (blue), DPPC + polyethylene-$b$-PEO (orange) and DPPC + 1,2 polybutadiene-$b$-PEO (purple) as well as the previously discussed DPPC + 1,4 polybutadiene system (pink).

The system of DPPC 1,4 PBD-$b$-PEO, which has been previously discussed in Figure \ref{fig:classification}, is shown again, now as a series of increasing concentration instead of sorted via classification.  When the PBD length is very short, at dilute concentrations, mixing is seen; the membrane thickness is similar to the pure lipid membrane, there is little variance in thickness, and the conditional entropy of mixing is relatively high. At higher polymer contents with a short polymer, the system forms polymer-rich and lipid-rich domains of similar thickness, maintaining a low variance of thickness but decreasing the entropy of mixing.
Longer polymers ($n_{PBD}>20$) unzip the membrane, resulting in a large variance of thickness, while low conditional entropy of mixing. Finally, polymer-rich bilayers have a low variance of thickness, low (but higher than unzipped/lateral phase separation) conditional entropy of mixing, and thickness similar to pure polymer membrane are seen. The thickness generally increases when some lipids are added as the lipids add molecules to the membrane while not substantially affecting membrane thickness.

The system of PE-$b$-PEO with DPPC represents a limiting case; lipid tails (C1 type beads) are chemically similar to the polyethylene monomers (C1 type beads). Without chemical immiscibility, the system still shows a preference for longer chains to exist in the midplane of the bilayer, and, unzipping is still seen, albeit to a lesser extent (Figure S15). With sufficiently long polymers there is a much lower conditional entropy of mixing compared to short polymers, indicating demixing (in the vertical direction). There is also a small peak in the variance of thickness with long polymers at dilute concentration, indicating unzipping. The thickness of the membranes themselves increase with increasing polymer content up until a point, then, the inclusion of additional polymer decreases the membrane thickness.

The system of 1,2 PBD-$b$-PEO + DPPC shows similar behavior to the polyethylene system. The parameterization for 1,2 PBD includes a small bead ($SC3$) and a tiny side-chain bead ($TC2$), each which is more `hydrophobic' than the 1,4 parameterization \cite{vreeker2025nanopore}. In addition to less chemical immiscibility, the use of two smaller beads favors mixing over demixing because of packing of differently sized spheres. Thus, although the entropy of mixing is lower than the pure system, and although the variance of thickness for the dilute, long polymer systems is higher, the system behaves more closely to PE rather than to DPPC + 1,4 polybutadiene. In Figure S16, it can be seen that while short polymers are correlated spatially, they are not clearly laterally phase separated. This aligns with the results of Vreeker and coworkers who needed a methylated lipid (DPhPC) \cite{vreeker2025nanopore} to observe lateral phase separation in the hybrid 1,2 PBD-$b$-PEO system. This also agrees with the results of Schlitter and coworkers \cite{schlitter2025nanoscale}, who only observe unzipping, polymer-rich, and mixing morphologies in a system with POPC and 1,2 PBD-$b$-PEO. One other model for 1,2 PBD-$b$-PEO does exist \cite{Muller2023} which uses tiny beads with reduced self-interactions. This parameterization always displayed mixing behavior due to the reduced self interactions, and is shown in Figures S17 and S18.

Systems containing 1,4 PBD-$b$-PEO + DOPC show behavior between DPPC + 1,4 PBD-b-PEO and DPPC 1,2 PBD-$b$-PEO due to a more fluid and less hydrophobic lipid being modeled. The reduced chemical immiscibility means that all short polymers remain mixed, with high conditional entropy of mixing and low variance of thickness. On the other hand, long polymers show very large unzipped domains, especially at low polymer content. Intermediate polymer lengths show a combination of mixing, lateral phase separation and unzipping (Figure S19).

Also simulated were DPPC molecules with slightly suppressed temperature (2.25 $k_BT$) which exist in the gel phase. Because of the arrested dynamics of the membranes, it is difficult to surmise the equilibrium properties of these phases. Notably, on the timescales of our simulations (as shown in Figure S11), hybrid membranes with gel phase lipids display faceted polymer-rich clusters consistent with $L_\alpha/L_\beta$ coexistence in lipid-lipid phase separation and can also display some unzipping-like behavior.

In summary, the four morphologies outlined in Section \ref{sec:morph} appear in multiple systems, not only in the presented system of DPPC and 1,4 PBD-$b$-PEO. Changing the components of the system changes the morphological behavior of the system in ways consistent with the physics of the proposed models (hydrophobicity differences, thickness mismatch), even in extreme cases (i.e. no hydrophobicity mismatch).  One particularly interesting feature is the disappearance of lateral phase separation, which, of the tested systems, only appears in 1,4 PBD-$b$-PEO and DPPC/DOPC. This is consistent with prior work which showed that more foreign lipids were needed to obtain lateral phase separation \cite{schlitter2025nanoscale}. To drive lateral phase separation, there needs to be minimal thickness mismatch and a substantial hydrophobic mismatch; increasing the thickness of the lipid membrane and using a less hydrophobic, short polymer can make a system undergo lateral phase separation. Investigating how cholesterol can mediate the phase separation of hybrid systems, enabling highly tuned membrane behavior, could be an exciting direction of future research. 

\section{Conclusion}
The morphology of hybrid lipid--block copolymer membranes can be described by the hydrophobic thickness mismatch, chemical immiscibility, and geometric constraints. Here, we demonstrated that the model derived by Wang and Safran for the thickness of amphiphilic monolayers \cite{wang1991curvature} also applies to amphiphilic bilayers. We extensively validated the influence of these effects on morphology using coarse-grained molecular dynamics simulations across several lipid types, polymer models, PBD lengths, PEO lengths and concentrations of polymer. 

This physical understanding is applicable to many system specific behaviors reported in the literature\cite{Go2021,le2013hybrid}, highlighting that the same physics govern most hybrid systems. The descriptors of relevance include the lipid membrane thickness, polymer membrane thickness, concentration, and chemical immiscibility. 

The \unit{\nm}-scale behaviors also define the \unit{\micro \m}-scale behavior. For instance, lateral phase separation should continue to macro-phase separation (when not in competition with hydrophilic brush repulsion). The $nm$-scale behavior also dictates the resulting properties of the membrane. Thus, this work enables rational design of hybrid membranes.

Finally, the topic of thermodynamically stable $nm$-sized domains is an area of further interest. An ultra coarse-grained model would enable investigations of longer time scales, larger length scales and faster dynamics to appropriately study several mechanisms to stabilize $nm$ sized domains. Small, $nm$ sized domains are uniquely advantageous for maximizing the synergistic mechanical properties of polymer membranes \cite{discher1999polymersomes} with the bio-compatibility of lipid membranes\cite{vreeker2025nanopore}. These features can mimic the functionality of lipid rafts seen in lipid membranes \cite{sezgin2017mystery} and expand our understanding of how cells create functional membranes. 

\section{Methods}
\subsection{System Preparation}
Membranes were initialized in a near-membrane state with lipids and polymers distributed randomly unless otherwise noted. Initial high energy configurations were removed following a brief FIRE energy minimization. The initialization and equilibration protocol is shown in Figure S7.

\subsection{Simulation Methods}

We performed coarse-grained molecular dynamics simulations using HOOMD-blue \cite{Anderson2020} and parameterized with the Martini v3 force field \cite{Souza2021}.  Simulations were conducted in an $NPT$ ensemble where the pressure in the $z$ direction was maintained at $1$ \unit{atm} and the pressure in the $xy$--plane was held at $0$ \unit{atm}. Unless otherwise noted, the $x$ and $y$ degrees of freedom were coupled. The temperature was maintained at $298$ \unit{\K}. A constant pressure barostat \cite{Martyna1994} and Nose-Hoover thermostat were used to maintain the pressure and temperature. A timestep of $20$ \unit{\fs} was used, consistent with Martini simulations. Electrostatics were handled with the reaction field scheme using a cutoff distance of $1.1$\unit{\nm}. Lennard-Jones type interactions also have $1.1$\unit{\nm} cutoff. All simulations were conducted with the simulation package HOOMD-blue \cite{Anderson2020}.

Generally, prior to the production simulation of a membrane, the following equilibration took place. A FIRE energy minimization \cite{Bitzek2006fire} was performed to remove initial overlap. FIRE was performed with a timestep of 0.001, a displacement capped integrator of 0.005\unit{\nm}, constant $N$ and constant $V$ ensemble. Bonded and angle interactions were taken from the Martini parameterization and a soft repulsive (DPD-type) interaction with $r_{cut}=0.5$\unit{\nm} was used, increasing the repulsion parameter, $A$, from 10 to 50 over 5 iterations of 100,000 steps. After every 100,000 timesteps (except the last), the water beads were integrated ($dt = 0.001 \unit{\ps}$) with a NVT thermostat (Bussi, 2.47kT) for 500 timesteps to provide sufficient thermal noise to prevent the water beads from crystallizing. Finally, prior to the main simulation run, a brief (1,000 timesteps, $dt = 0.002 \unit{\ps}$) was executed with all the main simulation parameters, including the appropriate LJ pair interactions. 

\subsubsection{Pure Polymer Simulations}

A complete table of all pure-polymer membrane simulations, including the final box dimensions and membrane thickness, is shown in Table S1. All membranes were sufficiently solvated such that the polymers extending from the top bilayer could not interact with the polymers extending from the bottom layers. The number of polymers and water content was not kept constant to account for the changing size of the polymers. 

Polymers were initialized in a helix conformation randomly in a bilayer-like configuration, with equal number of polymers pointing up and down. Sufficiently long polymers were initialized with some interdigitation. 
FIRE energy minimization and equilibration were performed to remove initial high energy configurations\cite{Bitzek2006fire}.

Simulations were run for 20,000,000 timesteps (0.4 \unit{\mu \s}), where their thicknesses were fully equilibrated (all trajectories are shown in Figure S1). Frames were saved every 100,000 timesteps.

\subsubsection{Hybrid Membrane Simulations}
Hybrid membranes were initialized randomly in a mixed bilayer-like configuration with polymers initially in helices and lipids nematically ordered in a bilayer. Membranes were solvated with sufficient water such that the hydrophilic polymers cannot interact across the periodic boundary condition. Because the membranes do not have constant polymer lengths, the size of the membranes varied. The average simulation corresponds to the volume of a 24 \unit{\nm} cube. The box, in the $x$ direction, ranged from 16 to 33 \unit{\nm}. In the $z$ dimension, the size ranged from 18 to 50 \unit{\nm}. The temperature was maintained at 2.47$k_BT$ unless otherwise noted (gel phase DPPC simulations). Frames were written every 100,000 timesteps.

As shown in Figure S8, membrane properties are reasonably equilibrated on the timescale of simulation $0.2$ \unit{\mu \s} with the exception of conditional entropy of mixing, which has a long tail due to the coarsening kinetics of domains. The coarsening of domains has little impact on the equilibrium structure of domains (as probed by the molecular dynamics length scale), and domain growth will be a subject of future work.

\subsubsection{Temperature Simulations}
Membranes were initialized with 50 mol \% polymer, where the polymer had 5 PBD monomers and 5 PEO monomers, randomly in a solvated bilayer-like configuration. After the simulation for 0.2 \unit{\mu \s} at 3.2$k_BT$, the dimensions of the simulation box were 27\unit{\nm} x 27\unit{\nm} x 14\unit{\nm}. There were 1008 lipids, 1008 polymers, and 55032 water beads in the simulation.

FIRE energy minimization and equilibration were performed to remove initial high energy configurations. Next, the membrane was simulated at a temperature of 3.2 $k_BT$ for 10,000,000 steps (0.2 \unit{\mu \s}).  The barostat and thermostat settings are identical to typical hybrid membrane simulations. Frames were written every 100,000 timesteps. This step equilibrates the membrane configuration at a high-temperature.

Then, the temperature cycling was performed. Again, all barostats and thermostat settings are the same as normal hybrid simulations. The temperature is instantaneously increased to 3.75 $k_BT$. For the cycling, frames are written every 200,000 timesteps. The temperature is changed in a stepwise manner, with 200 steps between the temperature of 3.75 $k_BT$ and 2.25 $k_BT$. At each step, the membrane is simulated for 500,000 timesteps. This cycling was repeated in the heating direction, and then cooling again, as well as for three uniquely initialized replicates. In total, over 18 \unit{\mu \s} of simulation is analyzed for this investigation. The heating rate corresponds to $0.75$ \unit{\K \per \mu \s}. There is minimal hysteresis on heating or cooling, indicating that the heating rate is reasonable for the observed properties on the length scales probed.
When the temperature is changed, both the temperature of the thermostat is updated and the velocities of all particles are changed to the set temperature, as drawn from a Boltzmann distribution.

Observed properties (area, 2D order parameter) are reported as smoothed by a convolution with a window size of 20 (corresponding to a temperature window of 0.019 $k_BT$).

\subsubsection{Interfacial Simulations}
A membrane was initialized with a box length of 68 \unit{\nm}, a width of 10\unit{\nm} and a height of 34 \unit{\nm}, containing 30 mol \% PBD$_{35}$-$b$-PEO$_{24}$ and DOPC as the lipid. This corresponded to 1254 lipids, 538 polymers, and 153542 water beads. The initial configuration was phase separated, as shown in Figure \ref{fig:interface}. FIRE energy minimization and equilibration were performed  to remove initial high energy configurations.

Then, the membrane was simulated for 10.52 \unit{\mu \s}, frames were written every 100,000 timestep. The barostat and thermostat are the same as the hybrid membranes, with the only difference being the $y$ dimension was removed as a degree of freedom for the barostat.
Temperature was maintained at 2.47 $k_BT$.

\subsection{Molecular Parameterization}
The PBD model was developed and verified in Ref.\!\citenum{Muller2023}. To match the experimentally used polymer, we used the trans 1,4 polybutadiene parameterization, which is the ``one bead'' parameterization from Ref. \citenum{Muller2023}. This parameterization consists of a chain of C4 beads, which are also used to model butane \cite{Alessandri2022}. We have previously validated the polymer size in various solvents \cite{Tallman2024}.

Variants of the PBD model were also used. First, to model a polymer with no hydrophobicity difference compared to lipid tails, the C4 bead was changed to a C1 bead. This effectively models a polyethylene-type monomer where each bead represents two polyethylene monomers. The bond and angle potentials were carried over from the PBD model.
Second, we also simulated systems with a 1,2 poly-butadiene. 
Two parameterizations exist in the literature: One parameterization uses a TC2r backbone bead and a TC4r side chain bead and is reported to favor mixing\cite{Muller2023}. The second uses a TC2 backbone bead and an SC3 side chain bead \cite{vreeker2025nanopore}. This shows lateral phase separation when mixed with a DPhPc lipid \cite{vreeker2025nanopore} and unzipped/polymer-rich morphologies with DOPC and POPC in the range of PBD$_{11}$-$b$-PEO$_8$\cite{vreeker2025nanopore,schlitter2025nanoscale}. Our implementation of the second model replaces the combined bending and torsion dihedral potential with a bond angle due to the limited dihedral potentials available in HOOMD-blue.

The PEO model was adopted from recent models developed for Martini v2 \cite{Rossi2012,Grunewald2018} and updated for Martini v3 \cite{polyply}. It should be noted that this model uses a combined bending and torsion angle to resolve a previously described numerical instability. We previously observed similar scaling exponents in water with and without this dihedral potential and therefore do not use the dihedral potential.
Some works have reduced the interaction strength between PEO SN3r-type beads and lipid head groups (Q1/Q5) to decrease the adsorption between PEO and the lipid membrane \cite{vreeker2025nanopore}. Muller and coworkers reduced the interaction energy between TC4r beads and PEO chain to similarly decrease adsorption and incorporation of PEO chains in the polymer membrane\cite{Muller2023}. We do not make these modifications but do not expect that it would have a significant influence on the results.

\subsection{Analysis}
\subsubsection{Extracting Parameters for Pure Polymer Membranes}
To obtain the parameters which determine the thickness of polymer membranes, we utilized several methods discussed in full in the supplementary information section I.A.1. We performed separate simulations of PBD in a polymer melt to determine characteristic ratio of PBD. The PEO brush density was obtained by fitting the PEO brushes from all membrane simulations to a Milner, Witten, and Cates parabolic distribution \cite{milner1988theory}. The PEO chain stiffness, the $xy$ chain stiffness (which enabled normalizing the grafting density to statistical segments), and the excluded volume were extracted. The PEO monomer size, and its corresponding volume, were obtained from the Lennard Jones and bonded parameters.  
The interfacial tension was the single free parameter obtained by fitting the model to the simulated thickness data. 

\subsubsection{Pure Polymer Membrane Thickness}
The thickness of the pure polymer membranes was computed by removing all water beads and PEO beads from the simulation frame. Then, a mesh was created over the membrane using an Ovito surface mesh \cite{Stukowski2009} with a alpha probe sphere radius of 0.75 \unit{\nm}. The two meshes for the upper and lower leaflet were separated and discretized onto a 100x100 grid, collecting the $z$ position for a given $xy$ point with a linear interpolation (using scipy griddata). The local thickness then was computed with the following formula:
\begin{equation}
    t = \frac{1}{n_x \cdot n_y}\sum_{i\in \{x,y\}} (z^u_i - z^l_i)\quad .
\end{equation}  

\subsubsection{Hybrid Membrane Thickness}
The thickness of the polymer membrane is computed via selecting all hydrophobic beads (excluding all water beads, as well as PEO beads for polymer membranes or $Q1$ and $Q5$ beads for lipid membranes). A grid with resolution of 1\unit{\nm} was constructed. For each grid point, all beads within 2\unit{\nm} were identified. The five beads with the lowest and the highest $z$ coordinate were averaged. The thickness of the membrane is then defined to be \begin{equation}
    t = \frac{1}{n_x \cdot n_y}\sum_{i\in \{x,y\}} (z^u_i - z^l_i)\quad .
\end{equation}  
The variance of thickness is defined as 
\begin{equation}
    Var\left(t\right) = \frac{1}{n_x \cdot n_y}\sum_{i\in \{x,y\}} (z^u_i - z^l_i)^2\quad .
\end{equation}  
The two thickness calculation methods give consistent results for the systems studied.

\subsubsection{Identifying Morphology Types}
The four morphologies have unique characteristics which make them identifiable. Polymer-rich morphologies have a continuous polymer segment which spans the entire simulation box, and lipids are displaced from the membrane midplane. Unzipped morphologies have some polymer segments but they do not span the entire simulation box. Additionally, they have some subset of lipids which are far from the membrane midplane. Mixed morphologies  have no clusters, whereas lateral phase separation has clusters of both lipids and polymers, but no lipids are displaced from the membrane midplane.

For DPPC + 1,4 PBD-$b$-PEO, we evaluated the morphologies both by visual assessment and also with properties of the simulations. The thresholds and properties of interest were determined by feeding many properties and the visually classified morphology into a decision tree classifier. This selected the three properties and thresholds that best classified the membranes according to hand-labeled classifications. The input properties were number of lipid clusters identified by DBSCAN\cite{DBScan} (min samples 20, eps=1.1\unit{\nm}), number of polymer clusters, standard deviation of lipid $xy$ histogram (a high value represents more phase separation), standard deviation of polymer $xy$ histogram, mean membrane thickness, standard deviation of membrane thickness (a high value indicates coexistence of thick and thin regions), area per lipid, percent of membrane area covered by lipids, percent of membrane area covered by polymers, and percent of lipids within 2.4 \unit{\nm} of the membrane surface. 
These quantities were calculated by averaging over the last 10 frames of each simulation.

The decision tree identified that the three parameters necessary to classify most of the data are standard deviation of polymer $xy$ histogram, percent of lipids within 2.4 \unit{\nm} of the surface and number of lipid clusters. This aligns closely with a naive guess of how to classify the system. It also correctly classifies 74/79 of the data points. The data points it does not classify correctly all appear in the transition region between phases, where the snapshots display characteristics of two or more morphologies. The final decision tree is shown in Figure S9.

For other types of lipids and polymers, classification was only done by visual inspection based on the first section, assessing if there was a continuous monolayer of lipids, coexistence of thick and thin phases, lateral phase separation or mixing.

\subsubsection{Area Per Molecule}
The area per molecule was calculated with average of the box area and the number of molecules: $APM = 2 l_x l_y/n_m$.

\subsubsection{Conditional Entropy of Mixing}
The normalized conditional entropy of mixing was calculated according to prior work by Brandani and coworkers \cite{brandani2013quantifying}.
The normalized conditional entropy of mixing was calculated per bead (and not per lipid), ignoring any beads directly bonded. Hydrophobic lipid beads (C1, C4h) were assigned type of 'L' and were considered identical. Hydrophobic polymer beads (C4, TC3, SC4, TC2r, TC4r) were assigned type 'P' and considered identical. The local environment of each bead was obtained from all non-excluded beads within a 3-D sphere of radius 0.85\unit{\nm} from the central bead. When presented, conditional entropy is averaged from the final two frames of simulation.

\subsubsection{Mixing Order Parameter}
The mixing order parameter, $\langle \psi_{xy} \rangle$, is used to calculate the average mixing of 2D system from two normalized density histogram (in this case, hydrophobic tails of lipids, $P^{l}_{xy}$ and hydrophobic polymers, $P^{p}_{xy}$) with the expression 
\begin{equation}
    \langle \psi_{xy} \rangle = 1- \frac{\langle P^{l}_{xy}P^{p}_{xy}\rangle }{\langle P^{p}_{xy}\rangle \langle P^{l}_{xy} \rangle} \quad,
\end{equation}
such that if the two histograms overlap, the order parameter is zero because $\langle P^{l}_{xy}P^{p}_{xy}\rangle=\langle P^{p}_{xy}\rangle \langle P^{l}_{xy} \rangle$, and if the two histograms have no overlap, the order parameter is unity. There are 40 bins in each direction, as shown in Figure \ref{fig:temp_effect}. 

The reason to use the 2D order parameter $\langle \psi_{xy} \rangle$ instead of the normalized conditional entropy of mixing is because the calculation is computationally cheaper, measures similar information, and is well established \cite{jeong2024modeling}. The normalized conditional entropy of mixing, as implemented, provides 3D information regarding demixing, whereas here, this order parameter only provides 2D information.

\subsubsection{Polymer Chain Properties}
Three measures of radius of gyration are reported in Figure \ref{fig:interface}. These are obtained by monitoring the radius of gyration in a particular dimension, averaged over 1,000 frames and binned corresponding to the center of mass of the particular chain. The 1D radius of gyration for PEO and PBD blocks is measured by calculating the average distance of all beads of type $i$ in a given dimension from the chain's center of mass \begin{equation}
    R_{\alpha,i} = \frac{1}{N} \sum_{j=1}^N \left( R_{\alpha, i}^j - R_{\alpha, i}^{com}\right)^2 \quad.
\end{equation}
The total radius of gyration (denoted $xyz$) is given by 
\begin{equation}
    R_{i} = \frac{1}{N} \sum_{j=1}^N \left| R_{i}^j - R_{i}^{com} \right|^2
\end{equation}

\subsubsection{Rendering}
All snapshots were rendered using the analysis and visualization software Ovito \cite{Stukowski2009}. Polymers are rendered in red and lipids in blue across all snapshots.

\section{Supporting Information}
Additional simulation results and additional discussions for all systems discussed in the main text are included in the supporting information.  

\section{Acknowledgments}
This work made use of the Illinois Campus Cluster, a computing resource that is operated by the Illinois Campus Cluster Program (ICCP) in conjunction with the National Center for Supercomputing Applications (NCSA) and which is supported by funds from UIUC.
The authors would like to acknowledge Charles Sing for useful conversations on scaling arguments of bilayer systems.
\bibliography{bib}
\end{document}

\setcitestyle{super}

\title{Supplementary Information: Phase Behavior of Unilamellar Hybrid Lipid-Diblock Copolymer Membranes}
\author{James F. Tallman}
\author{Junyu Wu}
 \author{Antonia Statt}
  \affiliation{Department of Materials Science and Engineering, The Grainger College of Engineering, University of Illinois Urbana-Champaign, Urbana, Illinois 61801, USA}
\maketitle

\tableofcontents 
\section{Pure Polymer Membranes}

\subsection{Simulation Details}

\begin{table}[]
    \centering
        \caption{Summary of all pure polymer membranes (defined by their PBD and PEO block lengths) simulated, as well as their recorded stability, dimensions in the $x$ and $z$ direction and equilibrated thicknesses}
    \begin{tabular}{ccccccc}
PBD Monomers & PEO Monomers & Stability & Water Beads  & $x$ dimension (\unit{\nm})  & $z$ dimension (\unit{\nm}) & Final Thickness (\unit{\nm})\\\hline

5 & 10 & Unstable & 108544 & NA & NA & NA \\
5 & 15 & Unstable & 107520 & NA & NA & NA \\
5 & 25 & Unstable & 104448 & NA & NA & NA \\
10 & 25 & Unstable & 102400 & NA & NA & NA \\
10 & 50 & Unstable & 116736 & NA & NA & NA \\
10 & 75 & Unstable & 141312 & NA & NA & NA \\
25 & 50 & Stable & 114688 & 24.3 & 32.4 & 4.3 \\
25 & 75 & Stable & 137216 & 26.6 & 32.8 & 3.5 \\
50 & 50 & Stable & 114688 & 25.3 & 33.9 & 7.9 \\
15 & 25 & Stable & 100352 & 20.7 & 34.9 & 3.5 \\
50 & 25 & Stable & 104448 & 22.6 & 37.2 & 9.8 \\
50 & 100 & Stable & 178176 & 28.9 & 38.4 & 5.9 \\
25 & 25 & Stable & 110592 & 21.1 & 38.7 & 5.7 \\
25 & 15 & Stable & 100352 & 19.2 & 41.9 & 6.9 \\
100 & 50 & Stable & 133120 & 26.2 & 42.1 & 14.6 \\
10 & 15 & Stable & 105472 & 18.5 & 42.6 & 2.9 \\
75 & 25 & Stable & 116736 & 23.3 & 42.6 & 13.9 \\
15 & 15 & Stable & 104448 & 18.6 & 43.4 & 4.4 \\
50 & 10 & Stable & 97281 & 19.8 & 44.6 & 12.9 \\
25 & 100 & Stable & 237568 & 28.6 & 44.7 & 3.0 \\
50 & 75 & Stable & 217088 & 27.4 & 47.0 & 6.7 \\
15 & 10 & Stable & 105472 & 17.6 & 47.9 & 4.9 \\
25 & 10 & Stable & 102400 & 17.9 & 48.2 & 7.9 \\
75 & 100 & Stable & 243712 & 29.3 & 49.3 & 8.8 \\
10 & 10 & Stable & 107520 & 17.2 & 49.4 & 3.5 \\
75 & 10 & Stable & 110592 & 20.8 & 49.7 & 17.4 \\
75 & 75 & Stable & 221184 & 27.8 & 49.8 & 9.9 \\
75 & 50 & Stable & 198656 & 25.9 & 51.4 & 11.3 \\
100 & 100 & Stable & 253954 & 29.6 & 52.7 & 11.5 \\
100 & 75 & Stable & 233472 & 28.1 & 53.9 & 12.9 \\
150 & 100 & Stable & 221184 & 29.5 & 54.5 & 17.3 \\
100 & 10 & Stable & 122880 & 21.5 & 55.7 & 21.9 \\
25 & 5 & Stable & 104448 & 16.4 & 57.2 & 9.6 \\
10 & 5 & Stable & 108544 & 15.4 & 60.6 & 4.3 \\
15 & 5 & Stable & 107521 & 15.6 & 60.9 & 6.3 \\
100 & 150 & Stable & 376834 & 32.0 & 62.3 & 9.8 \\
5 & 5 & Stable & 110592 & 15.0 & 63.4 & 2.2 \\\hline
    \end{tabular}
    \label{tab:all-pure-polym}
\end{table}

\begin{figure}
    \centering
    \includegraphics[width=.5\linewidth]{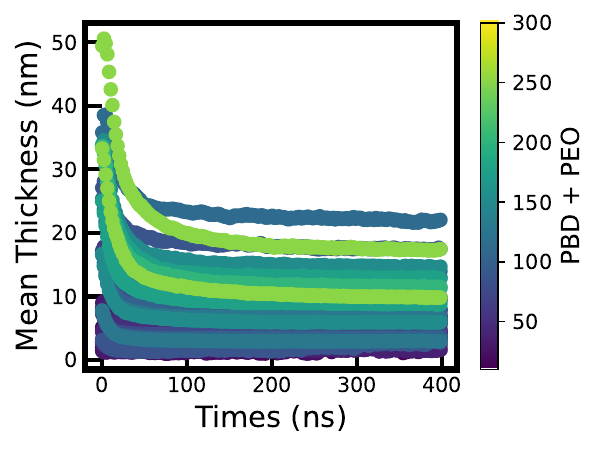}
    \caption{Equilibration of thickness of pure polymer membranes. Color indicates the total length of the polymer (in monomers). Simulations were conducted for 400ns.}
    \label{fig:equil-polym-thickness}
\end{figure}
\FloatBarrier

\subsubsection{Extracting Parameters from Simulations}
\begin{figure}
    \centering
    \includegraphics[width=0.49\linewidth]{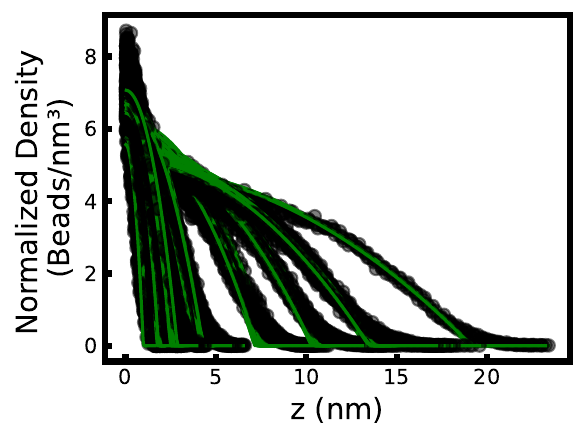}
    \includegraphics[width=0.49\linewidth]{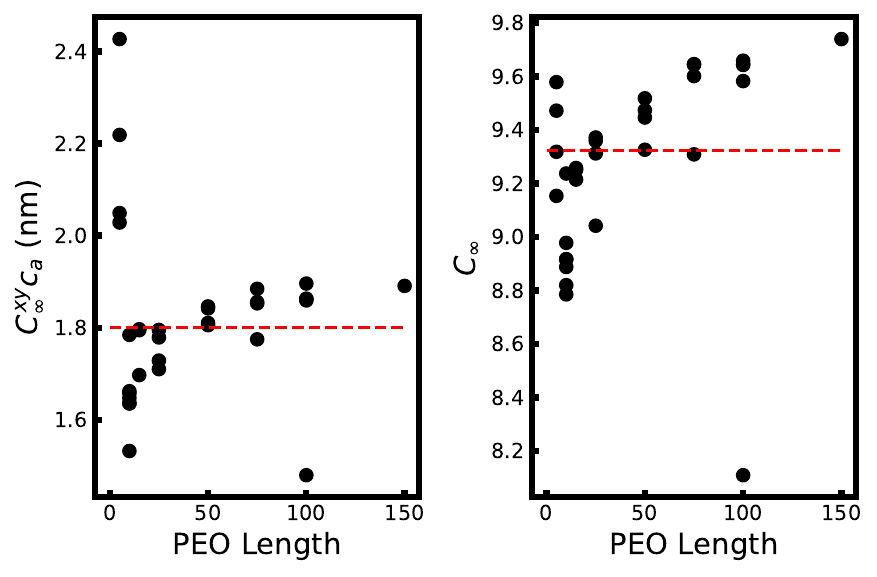}
    \caption{(Left) Fits of the normalized density as function of height $z$. All fits have an $R^2$ value greater than 0.9. Both bilayers are shifted such that $z=0$ is the point prior to the mode. Each curve is fit with a parabolic distribution. The last 15 \% of points in each series were removed because chain ends do not decay parabolically to zero. (Right) Average values of the joint parameter excluded volume times lateral segment size and the chain stiffness. The red lines indicates the median parameters which were selected for the resulting fit.}
    \label{fig:mwc-params}
\end{figure}

There are five parameters for the derived theory for bilayer thickness: PBD chain stiffness ($C_{\infty,b}$), PBD monomer size ($a_b$), ratio of the thermal energy to the interfacial tension ($\frac{k_BT}{\gamma}$),  PEO excluded volume ($c_a$), and PEO chain stiffness ($C_{\infty,a}$) and the anisotropic PEO chain stiffness ($C^{xy}_{\infty,a}$). We extract them from separate simulations or from the microscopic configurations of pure polymer simulations as follows:
\begin{description}
    \item[PBD chain stiffness] A polymer melt containing 120 PBD polymers with 50 monomers per polymer was simulated in an NPT ensemble (1 atm, coupled `xyz') for 20,000,000 timesteps. The average cosine of the angle between beads a distance of $j$ monomers apart was calculated. The characteristic ratio was calculated to be $1+2\cdot \sum_{j=1}^{50}\cos(\theta_j)$ \cite{rubinstein2023polymer} averaging over 50 frames (5,000,000 timesteps) and all polymers in each frame. The calculated characteristic ratio for PBD in simulation was 3.58 monomers.
    \item[PBD Monomer Volume] The volume of PBD monomers was calculated from pure-polymer membranes to be $a^3_b = \frac{dA}{n}$ where d is the membrane thickness, $A$ is the membrane area and $n$ is the number of PBD beads. This was averaged over several frames and trajectories and calculated to be $a_b=0.458$nm.
    \item[Interfacial tension ratio $\frac{kT}{\gamma}$] The thermal energy is maintained by a Nose Hoover thermostat to 2.47 kJ/mol. The interfacial tension proved difficult to approximate and instead was fit with the rest of the parameters fixed to the simulated thickness data. A final interfacial tension was fitted to be $11.7$ \unit{\kJ \per \mol}.
    \item[PEO Excluded Volume] We assume the density of the PEO corona to be a polymer brush (as shown  experimentally)\cite{smart2009polymersomes}. From this assumption, we fit the distribution of all PEO chains to a parabolic distribution as defined by MWC theory \cite{milner1988theory}.
    Because we ultimately want to discuss monomers and not statistical segments, we convert between them with a characteristic ratio. 
    
    One peculiarity is the definition of the dimensionless surface coverage in MWC's theory. The Kuhn segment length is assumed to be isotropic. However, the surface coverage is describing the in-plane concentration of chain ends at the interface. Thus, the in plane statistical segment size of the polymer should define this renormalization length, not the Kuhn segment. For that reason, we decouple these components and have two separate stiffness variables for PEO segments,  $C_{\infty,a}$ sets the Kuhn segment size and $C^{xy}_{\infty,a}$ sets the in plane segment size.

    The expressions which defines the concentration of polymers a distance $z$ from the surface in a polymer brush is 
    \begin{align}
        N &= n_{PEO}/C_{\infty,a} \\
        \tilde \sigma &= \sigma b_{a,xy}^2 =\sigma (a_a C^{xy}_{\infty,a})^2\\ 
        \tilde v_a &= \frac{c_a^2}{\left(a_a\right)^2} \\
        h^\star &= \left(\frac{12}{\pi^2}\right)^{1/3} \left( \tilde \sigma \tilde v_a \right)^{1/3}N\\
        B &= \frac{\pi^2}{8N^2}\\
        \phi(z) &= \frac{B}{\tilde v}\left( (h^\star)^2 - z^2 \right)C_{\infty,a} \quad, 
    \end{align}
    where $b_a$ is the size of the Kuhn segment, $c_a$ is the radius of the excluded volume cylinder, and $a_a$ is the PEO monomer size, $C_{\infty,a}$ is the characteristic ratio of the chain (isotropic), and $C_{\infty,a}^{xy}$ is the characteristic ratio in the $xy$ dimension. We combine the product $c_a\cdot C^{xy}_{\infty,a}$ in fitting because they are degenerate. $\phi(z)$ reports the concentration in units of monomers per nanometer.
    
    These equations are fit for the parameters $C^{xy}_{\infty,a}\cdot c_a$ and $C_{\infty,a}$. The resulting fits are shown in Figure \ref{fig:mwc-params} and the resulting parameters as shown as a function of PEO length are shown in Figure \ref{fig:mwc-params}.
    From this, we can extract that the average product of the excluded volume and chain stiffness, $c_a\cdot C^{xy}_{\infty,a}$, to be $1.8\,$nm. 
    
    \item[PEO Chain stiffness] From the same analysis, the PEO chain stiffness was extracted to be 9.39. Interestingly, this agrees with the measured characteristic ratio obtained from measuring the correlation of bond angles in PEO in a polymer brush (9.1). Both these values are significantly higher than the stiffness of solvated PEO homo-polymers in dilute solution (4.0). 
\end{description}

The values of all parameters used in the free energy expression and their origin are written in Table \ref{tab:all_param}
\begin{table}[]
    \centering
        \caption{Parameters used in the free energy expression (Eq. (1) of the main text), extracted from either fitting the polymer brush distribution ($C_{\infty,PEO}$, $C_{\infty,PEO,xy} \cdot c_a $) measuring from bulk simulations ($C_{\infty,PBD}$, $a_b$), taken from the Martini parameterization ($kT$) or fitted to match the thickness data ($\gamma$).}
    \begin{tabular}{cc} \hline 
         Parameter & Value  \\\hline 
         $C_{\infty,PBD}$& 3.54 \\
         $C_{\infty,PEO}$& 9.02 \\
         $C_{\infty,PEO,xy} \cdot c_a $& 1.8 \unit{\nm} \\
         $\gamma$& 11.73 \unit{\kJ \per \K \mol \per \nm \squared}\\ 
         $kT$ & 2.47 \unit{\kJ \per \K \mol} \\ 
         $a_b$ & 0.486 \unit{\nm}\\ 
    \end{tabular}
    \label{tab:all_param}
\end{table}

\FloatBarrier
\subsubsection{2D Thickness Map}

\begin{figure}
    \centering
    \includegraphics[width=1\linewidth]{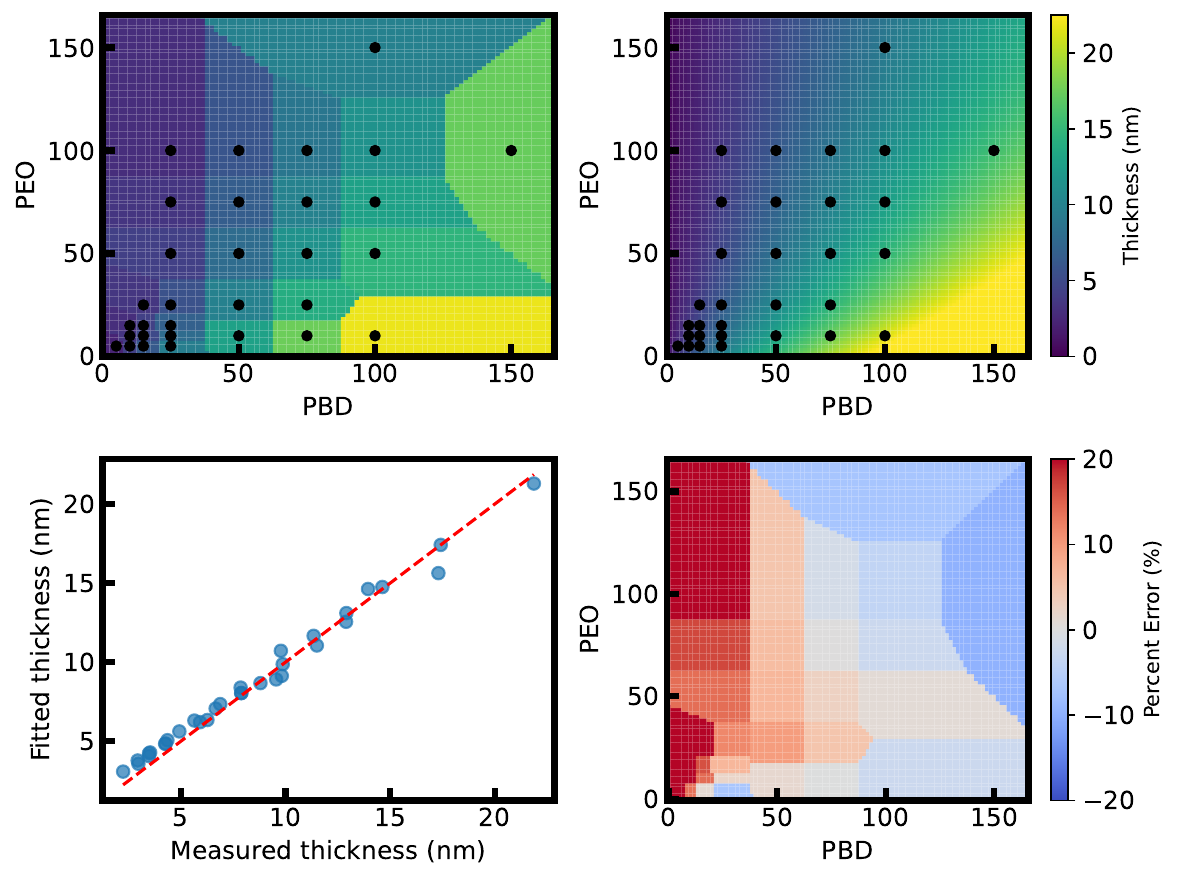}
    \caption{Comparison of the measured simulation thicknesses (top left) against the continuous thickness predictions (top right). Points on these plots reflect the points sampled in simulation. (Bottom left) parity plot showing strong agreement between the measured thickness and fitted thickness. (Bottom right) Percent error of thickness predictions from the actual measurements.  }
    \label{fig:2d_map}
\end{figure}
Figure  \ref{fig:2d_map} shows comparisons between theoretically predicted and measured thickness maps. 
Notably, the region of inaccurate predictions is at short PBD lengths and long PEO lengths, a region which is not likely to be accurately captured by the model because the large deformations induced by the polymer brush make the membrane very thin, which then breaks assumptions about the statistical nature of Kuhn segments (and also often leads to unstable membranes). 

\FloatBarrier
\subsubsection{Alignment of PEO and PBD Chains}
\begin{figure}
    \centering
    \includegraphics[width=0.75\linewidth]{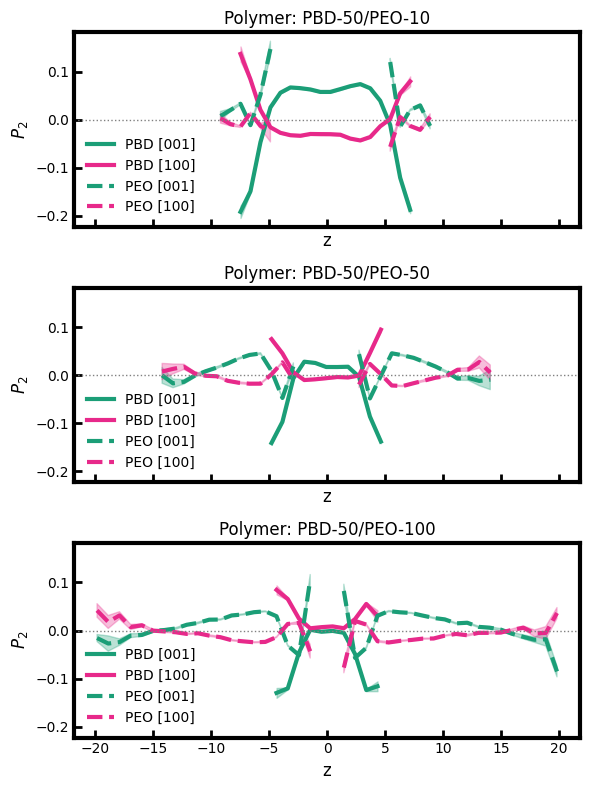}
    \caption{Alignment of PBD (solid lines) and PEO (dashed lines) chains with respect to the membrane normal (pink, [100]) and parallel (green [001]), shown for three membranes of increasing PEO chain length (decreasing thickness). }
    \label{fig:p2-order}
\end{figure}

Another point we can consider is the alignment of the polymer chains with varying PEO length (Figure \ref{fig:p2-order}). To analyze this, we use the P2 order parameter which is unity if bonds are aligned with a direction and -0.5 if they are perpendicular to that direction. This order parameter is computed as a function of the $z$ coordinate, where all bonds in a given bin (bin size = 1nm) over several frames are averaged. 

With a long PBD segment (50) and a short PEO segment (10), the PBD segment is aligned in the direction of the membrane normal in the bulk. At the interface with the PEO chains, the PBD chains are anti-aligned, running horizontally to the surface. The PEO chains are strongly aligned with the membrane normal close to the interface, however that order quickly decays and the chain ends are relatively uncorrelated.

With a longer PEO segment (50), the hydrophobic core of the membrane is thinned. This stretches the chains inside of the hydrophobic core, and thus they are less aligned to the membrane normal. The chains at the interface are still perpendicular to the membrane normal. The PEO chains are aligned to the direction of the membrane normal except very close to the interface (due to effects of the PEO-PBD bond) and at the chain end). The extent to which the PEO chains are aligned to the membrane normal decays roughly linearly with distance from the hydrophobic core. 

With an even longer PEO segment (100), these effects continue. Now, the membrane core is not aligned with the membrane normal direction at all, indicating that the chains are confined in the membrane, not stretched (as predicted by the strong segregation limit). The PEO segment is aligned to the membrane normal.

\subsubsection{Effective Scaling Exponents}
\begin{figure}
    \centering
    \includegraphics[width=0.7\linewidth]{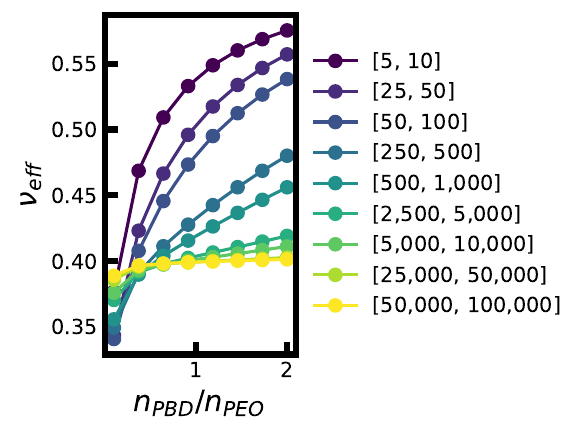}
    \caption{Effective scaling exponent as a function of ratio of $n_{PBD}/n_{PEO}$. The four series represent effective scaling exponents fit over different ranges of $n_{PBD}+n_{PEO}$.}
    \label{fig:power_law}
\end{figure}

 Because it is the competition of four energetic terms, a true scaling law is not expected to hold. To illustrate this point, in Figure \ref{fig:power_law}, the effective scaling exponents were calculated at fixed $n_{PBD}/n_{PEO}$ by fitting ranges of $n_{PBD}+n_{PEO}$ to a power law. This is then plotted for different ranges and ratios. 
When $n_{PEO} \gg n_{PBD}$, we do not expect the bilayer configuration to be even metastable. The effective scaling exponent in the limit as $n_{PBD}/n_{PEO}\rightarrow0$ is $\sim 0.25$. As $n_{PBD}/n_{PEO}\rightarrow\infty$, the effective scaling exponent goes to 0.66, recovering the strong segregation limit. At the intermediate values, the effective scaling exponent is strongly dependent on the length of the BCP. When the ratio of the monomers is one and the range is between 25 and 100 monomers (similar to most experimental systems, the effective scaling exponent is 0.5, consistent with the published literature.
\FloatBarrier
\subsubsection{Mechanical Properties}
\begin{figure}
    \centering
    \includegraphics[width=0.49\linewidth]{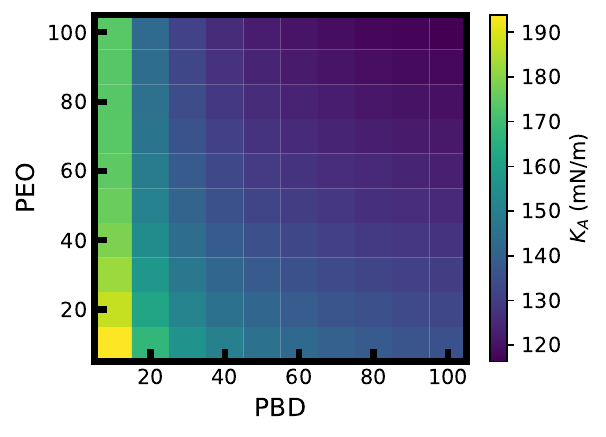} \includegraphics[width=0.49\linewidth]{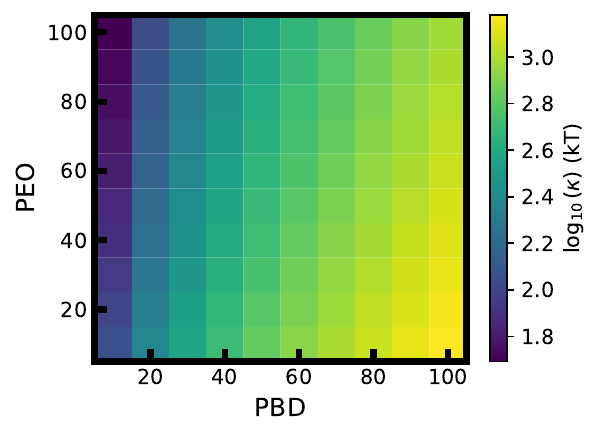}
    \caption{Mechanical properties of hybrid membranes from the free energy expression (Left) Area expansivity modulus shown as a function of PBD and PEO. (Right) Bending moduli shown as a function of PBD and PEO and the color bar is shown on a log scale.}
    \label{fig:mechprop}
\end{figure}

We can also extract mechanical properties from the free energy expression. \begin{equation}
    K_a = 2\frac{\partial^2 F}{\partial u^2 }\sigma
\end{equation}
where the factor of 2 originates as the free energy expression is an energy per chain and there are 2 monolayers.  In Fig. \ref{fig:mechprop}, we show the stretching moduli.  Stretching moduli are between 125 and 200 mN/m, similar to experimental energy scales. There is not a strong dependence of chain length when both PBD and PEO are increased simultaneously. The stiffest membranes are predicted to be at the pure PEO (very thin) or pure PBD (very thick) limits. The predicted stretching modulus and trends are similar to the experimentally observed measurements \cite{bermudez2002molecular}. 

The bending modulus (to the first order) can be expressed as 
\begin{equation}
    \kappa= \frac{1}{12}K_a d^2 \quad,
\end{equation}
since polymers are highly viscoelastic, bending is likely not to give rise to the typical tension and compression distribution. Rather, chains can relax to the new curvature being imposed. Nevertheless, the predicted bending moduli largely follows with the thickness of the membrane. Increasing the PBD length and PEO length simultaneously decreases the stretching moduli while it increases the thickness, leading to effectively no changes to the bending moduli.
\FloatBarrier

\section{Hybrid Membranes}
In this section, the additional details, results, and interpretations from hybrid membrane simulations are included and discussed. The results are grouped by the specific system (lipid and polymer type) under investigation.
\subsection{DPPC and PBD-$b$-PEO}
\subsubsection{Initialization and Equilibration of Hybrid Membranes}
\begin{figure}
    \centering
    \includegraphics[width=0.7\linewidth]{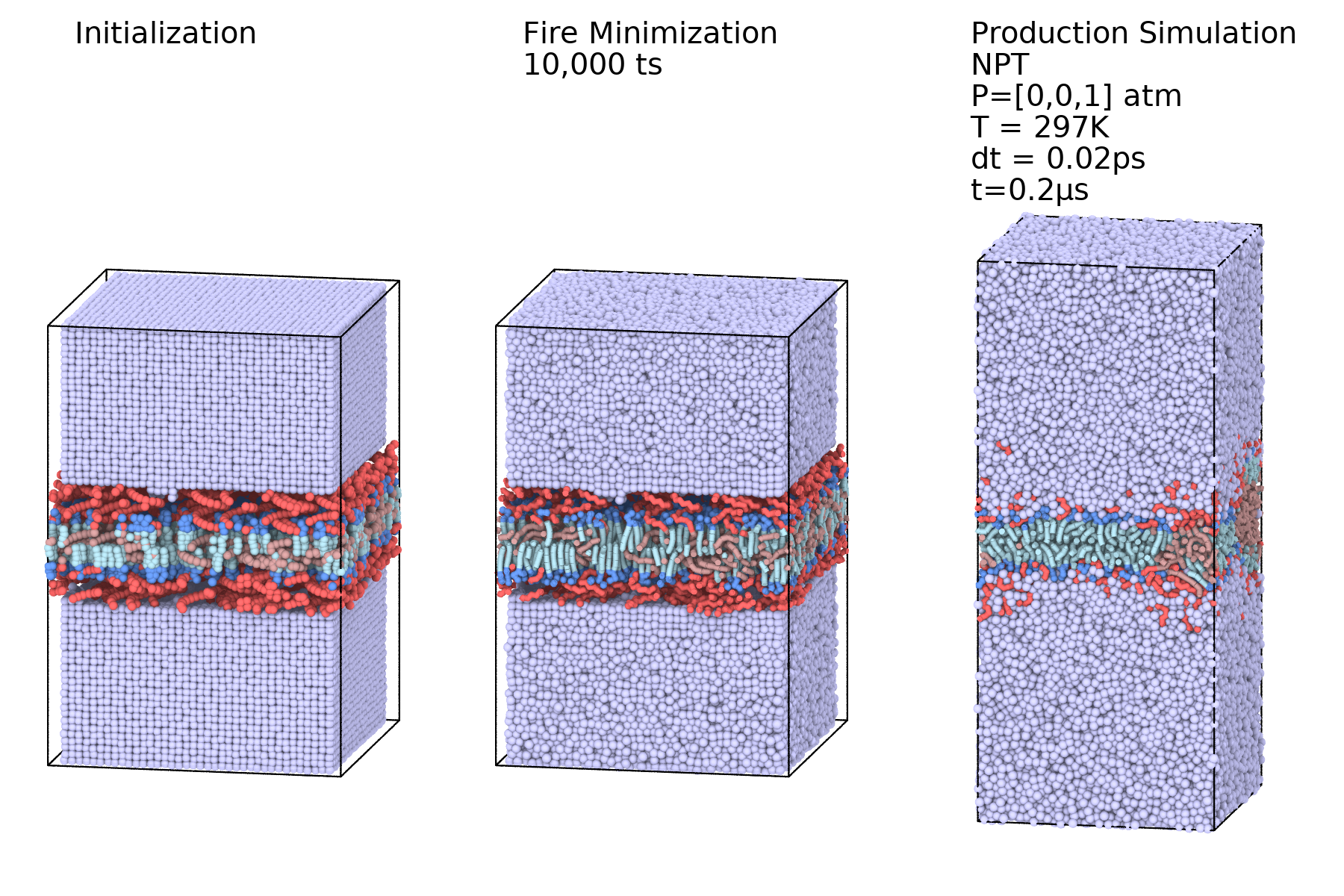}
    \caption{Initialization, equilibration and simulation protocol for hybrid membranes, where light blue/dark blue represent the lipid tail/head and light red/dark red represent the polymer hydrophobic tail/hydrophilic head. Water is shown in a light purple color. (Left) Morphology of membrane after initialization in a random bilayer-like configuration. (Center) Simulation after FIRE energy minimization used to remove initial overlaps. (Right) Simulation after NPT simulation for 0.2 $\mu s$.}
    \label{fig:init-and-equil}
\end{figure}

As discussed in the methods section, membranes are initialized in a membrane like configuration, overlaps are removed with a brief FIRE energy minimization (with a displacement capped integrator in an constant volume ensemble)\cite{Bitzek2006fire}, and then the simulation is conducted in a constant pressure ensemble. In Figure \ref{fig:init-and-equil}, this protocol is shown, with the first snapshot being after initialization, the second snapshot being after the FIRE energy minimization and the third snapshot being after the entire simulation is completed. 

\subsubsection{Equilibration of Hybrid Membrane Properties}
\begin{figure}
    \centering
    \includegraphics[width=0.45\linewidth]{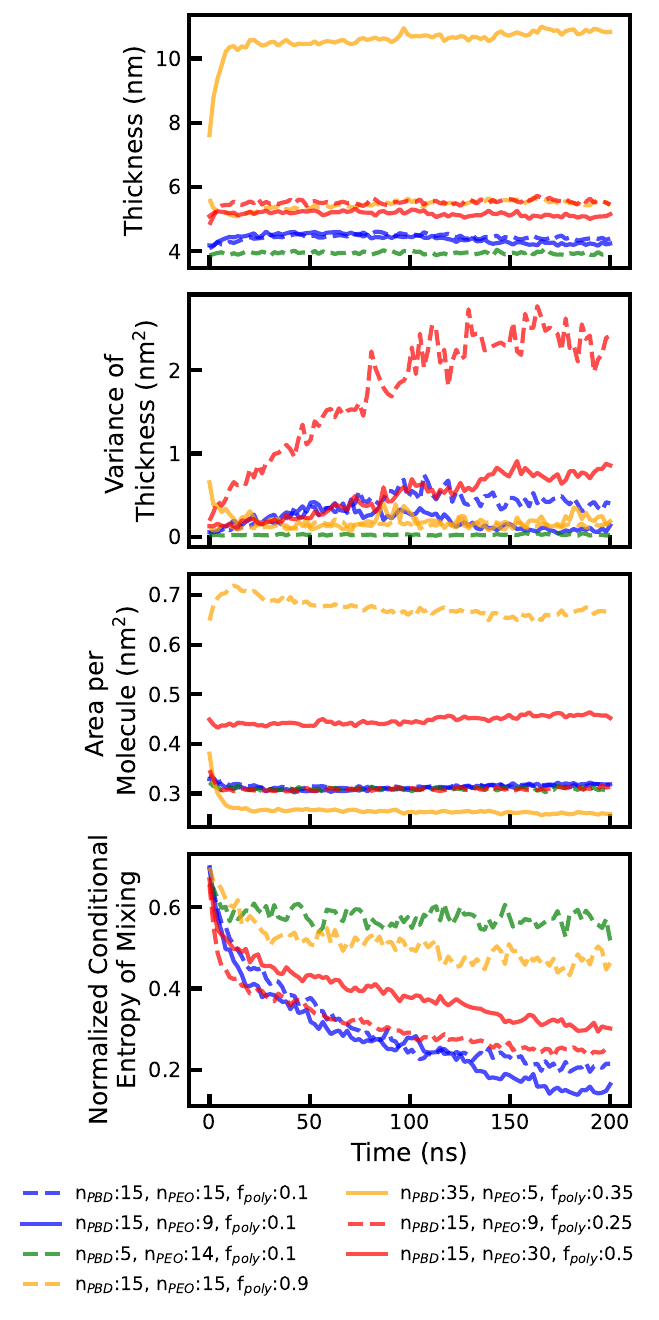}
    \caption{Equilibration of system properties over the simulation time for simulations representative of different distinct morphologies; mixing: green, lateral phase separation: blue, polymer-rich: yellow, unzipped: red. (1st) Thickness of the hydrophobic core of the membrane, including lipid tails and hydrophobic polymers. (2nd) Spatial variance of thickness in a single simulation frame, where a high value indicates non-constant thickness. (3rd) Area per molecule per monolayer. (4th) Normalized conditional entropy of mixing.}
    \label{fig:equil_times}
\end{figure}

The hybrid membranes were simulated for $0.2\mu$s. To determine if this simulation time was sufficient to discuss the thermodynamic states of hybrid membranes, we plot the different global membrane properties as a function of time for several hybrid membranes, as shown in Figure \ref{fig:equil_times}.

All membrane properties are reasonably equilibrated over the simulation time, such that the morphology or the ability to discuss said morphology will not change. The variance of thickness (averaged over last 10 frames) and conditional entropy of mixing (averaged over last 2 frames) have a long-timescale relaxation for the phase separating systems because of the slow process of domain fusion and coarsening --- this will be the focus of future work.

\FloatBarrier
\subsubsection{Lines on the Phase Map}
The parameters used for the construction of the Flory Huggins spinodal, as well as the reference area per lipid and DPPC membrane thickness used to draw the other lines on the phase map are summarized in Table \ref{tab:fh-chi}.
\begin{table}[]
    \centering
        \caption{Parameters used to draw the lines of the DPPC + 1,4 PBD-$b$-PEO at 2.47 kT phase map. Thickness and area per lipid are obtained from simulation of pure lipid membranes. Interaction parameters are taken from the potential well depth, $\epsilon$, of the polymer and lipid beads in the Martini model. The coordination number is the product of the height of the lipid tails (4 beads) and the average 2D coordination of spheres ($\sim 6$). Temperature was set from the simulation. The $\chi$ parameter is calculated from the above parameters, see eq. \ref{eq:chi}. The number of 2D monomers per lipid was 2, the number of 2D monomers per polymer was $n_p/4$.}
    \begin{tabular}{cc}\hline
        Parameter & Value \\\hline
         $t_{DPPC} $ & 4.07nm  \\ 
         $APL_{DPPC}$& 0.58 nm$^2$\\ 
         $w_{lipid-lipid}$& 3.39 kJ/K / mol \\
         $w_{polymer-polymer}$& 3.39 kJ/K / mol \\ 
         $w_{polymer-polymer}$& 3.07 kJ/K / mol \\
         $\#_{coord} = c\cdot z$& $6\cdot 4 =24$ \\ 
         $kT$& 2.47 kJ/K/mol \\ 
         $\chi$& 3.11 \\ 
         $N_l $ & 2 \\ 
         $N_p$ & $n_p/4$ \\ \hline
    \end{tabular}
    \label{tab:fh-chi}
\end{table}

The $\chi$ parameter is calculated with the equation: 
\begin{equation}
\label{eq:chi}
    \chi = \frac{c\cdot z}{kT} \cdot \left(\frac{1}{2}\left(w_{p-p} + w_{l-l}) - w_{p-l}\right)\right) \quad .
\end{equation}

The spinodal was constructed with obtaining the roots of the second derivative of the Flory-Huggins free energy\cite{rubinstein2023polymer}: \begin{equation}
    \frac{\partial F}{\partial \phi} =  \frac{1}{N_p  \phi} + \frac{1}{N_{l}  (1-\phi)} - 2\chi = 0 \quad .
\end{equation}

\subsubsection{Classification Procedure}

\begin{figure}
    \centering
    \includegraphics[width=0.95\linewidth]{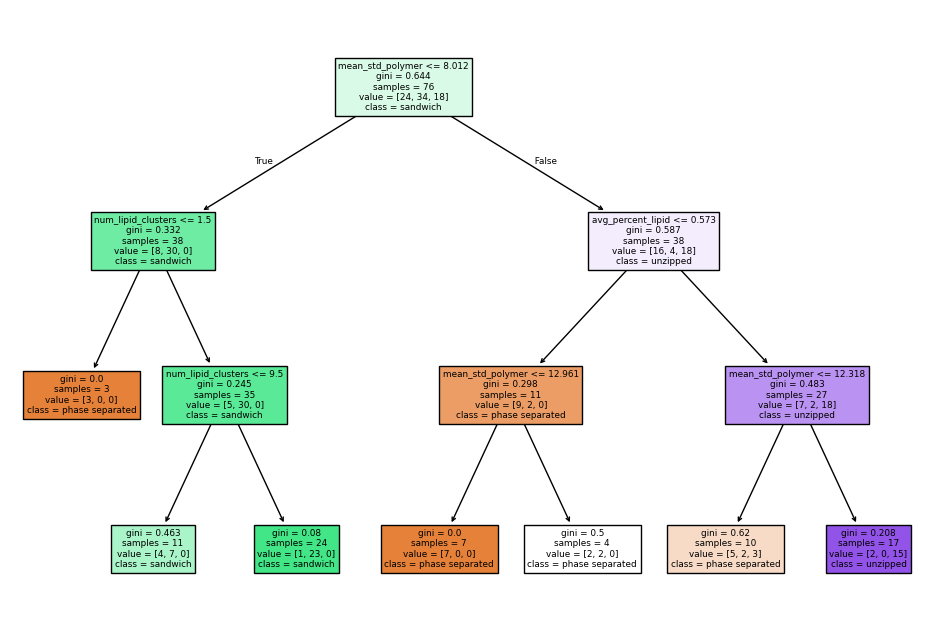}
    \caption{Decision tree computed from standard deviation of polymer $xy$ histogram, percent of lipids with 2.4 nm of the membrane midplane and number of lipid clusters}
    \label{fig:dec-tree}
\end{figure}

For all DPPC and PBD-$b$-PEO, all simulations were classified by an automated method and by manual identification. The automated method was devised by feeding many parameters into a decision tree classifier and identifying the three parameters and their threshold values that best classified the membranes according to the manually labeled classifications. As expected, the parameters the decision tree identified were the standard deviation of polymer $xy$ histogram, percent of lipids within 2.4 \unit{nm} of the midplane, and number of lipid clusters, which correctly classified 74/79 of the data points. The final decision tree with threshold values is shown in Figure \ref{fig:dec-tree}.

The selection of parameters was obtained by creating a decision tree with max depth of three (using \texttt{SKLearn.tree} decision tree classifier) for each set of three parameters from the eleven input parameters. The list of input parameters is provided in the methods. Using a test split size of 20\%, the best decision tree was selected by choosing the one which had the lowest error .

The other classification detail not encoded in the decision tree was if there were no lipid clusters or polymer clusters with a size of at least 20 beads, the morphology is counted as mixed. Only one mixed morphology was obtained within the tested parameters, hence why it was treated separately and identified manually.

\subsubsection{PBD and PEO Effect Hybrid Membrane Phase Behavior}

In the main text, we show the phase map with $n_{PBD}$ as the $y$ axis and concentration of polymer as the $x$ axis, as this correlates strongest with the resulting morphologies. However, this is truly a 3D phase map which we have projected into 2D, because the length of the PEO block also has an impact on the resulting structure. As the length of PEO chains are increased, they have an increased thinning effect on the membrane, preferring mixed or laterally phase separated morphologies.

\begin{figure}
    \centering
    \includegraphics[width=0.6\linewidth]{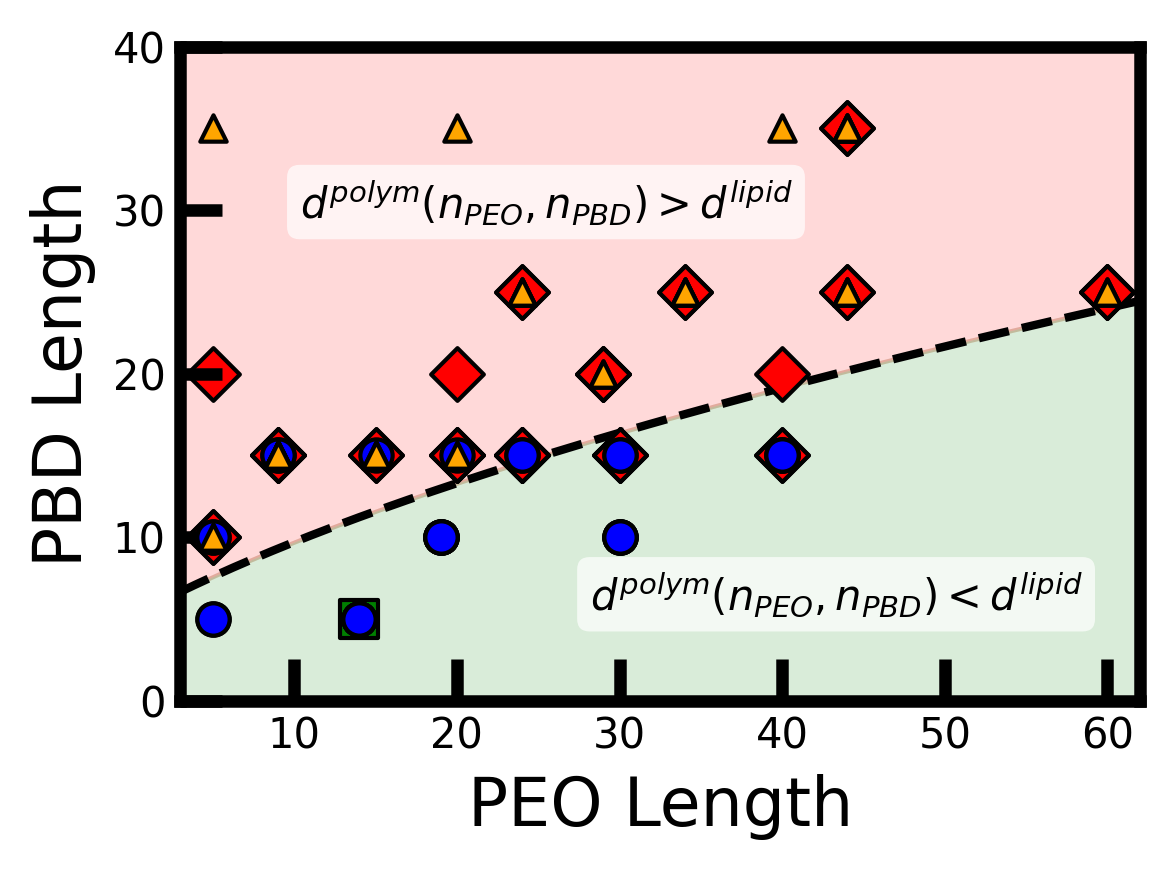}
    \caption{Phase map with concentration projected out, displaying PEO length vs. PBD length. Dotted line references the chain lengths at which the pure polymer membrane thickness is equivalent to the pure lipid membrane thickness. Symbols (polymer-rich: yellow triangle, unzipped: red rhombus, lateral phase separation: blue circle, mixing: green square) reflect the morphology of simulations that were performed with the given chain lengths.}
    \label{fig:pbd-peo-len}
\end{figure}

This trend is shown on the phase map in Figure \ref{fig:pbd-peo-len}. Here, the dotted line reflects the PBD/PEO length when the thickness of the pure polymer membrane is greater than that of the pure lipid membrane from the predicted theory. This prediction holds well, with some deviation at small PBD lengths where the thickness model does not hold as strongly and the morphologies are harder to distinguish.

Notably, it is impossible to separate between lateral phase separation and mixing or between unzipping and polymer-rich morphologies because the concentration axis of this diagram has been projected out. 

\subsubsection{Gel Phase DPPC and PBD-$b$-PEO hybrids }
\begin{figure}
    \centering
    \includegraphics[width=0.5\linewidth]{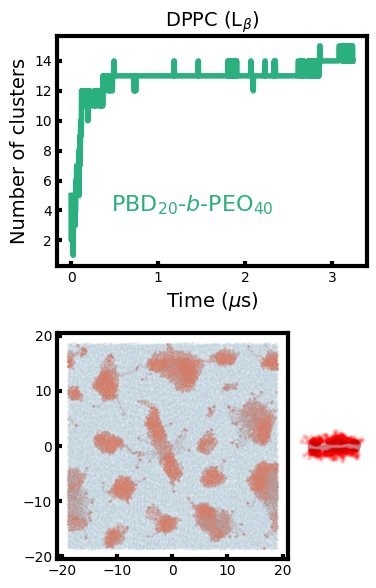}
    \caption{Simulation of DPPC at 2.25 $k_BT$, where DPPC is in the gel phase, the PBD length 20, PEO length 40, and 10 mol\% polymer is used. (Top) Number of clusters as a function of time, where simulation is conducted for 3 $\mu s$. (Bottom) Scatterplot of lipid and polymer positions from the final frame of simulation. Polymers are plotted with transparent red circles, lipids are plotted with transparent blue circles. (Bottom Right) Schematic of a single polymer domain when viewed from the side, where PEO beads are shown in dark red and PBD beads are pink.}
    \label{fig:gel-hybrid}
\end{figure}

Experimentally, DPPC is in the gel phase at room temperature. The Martini 3.0 parameterization of DPPC at room temperature, however, is in the liquid disordered phase. Reducing the temperature of the system from 2.47 $k_BT$ to 2.25 $k_BT$ causes the model DPPC lipid to transition into the gel phase. To investigate at a cursory level how the gel phase effects polymer structure, we simulated a system with PBD length 20, PEO length 40, and 10 mol\% polymer in the gel phase for 3 microseconds. Reported in Figure \ref{fig:gel-hybrid} is the result of the simulation. 

The dynamics of this system are significantly arrested. Polymer domains have long, faceted domains, do not coalesce, and polymer chains can be `stuck' in the lipid matrix. Additionally, both unzipped and laterally phase separated domains can be observed. This highlights that gel phase lipids are much more difficult to simulate because of arrested dynamics, but could be a system which stabilize finite-size domains over macroscopic phase separation.

\subsubsection{Temperature Effects at Varying Concentration}
\begin{figure}
    \centering
    \includegraphics[width=1\linewidth]{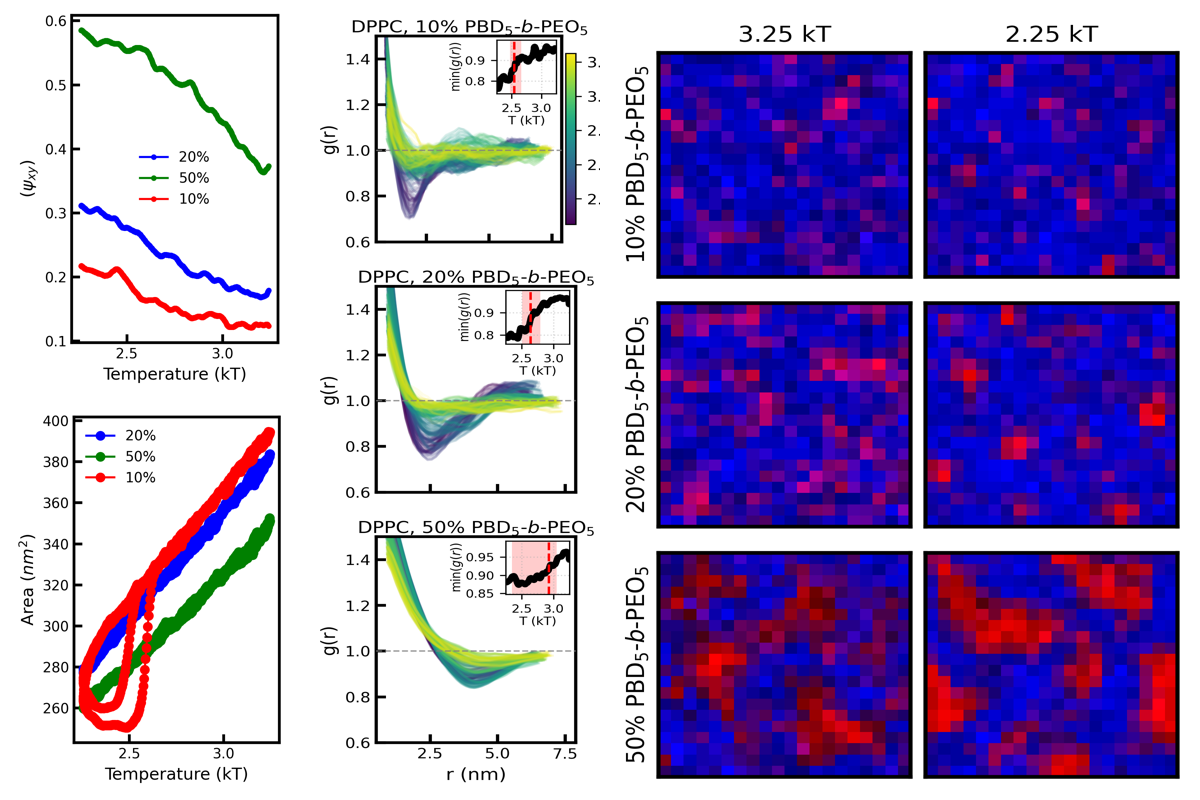}
    \caption{Investigation of temperature effect on phase separation of hybrid membranes with varying polymer content and PBD$_5$-$b$-PEO$_5$. (Top Left) Phase separation order parameter (0 if perfectly mixed, 1 if perfectly phase separated) plotted as a function of temperature. Order parameter is smoothed via convolution with window size of 20 frames. Order parameter averaged over heating and cooling. Color indicates the concentration of polymer. (Bottom left) Area plotted as a function of temperature using the above color scheme. Color indicates the concentration of polymer. (Center) Pair correlation functions between polymer species of (top) 10 mol\% polymer (middle) 20 mol\% polymer (bottom) 50 mol\% polymer. Green lines indicate values taken from high temperature, and purple lines indicate correlation functions from the lowest temperatures. The Insets are plots of the minimum value of the pair correlation function. (Right) Histograms show positions of lipid tails (blue) and hydrophobic polymer blocks (red) at the low temperature (right) and high temperature (left) extremes. }
    \label{fig:temp_conc}
\end{figure}

In addition to the length temperature sweep discussed in the main text at 50 mol \% polymer, two additional simulations were performed at 10 and 20 mol \% and are summarized in Figure \ref{fig:temp_conc}. Here, it can be seen that the order parameter does not display as large of a variance at the lower concentrations. This is partially due to the bin size, which would view small domains as mixed. Additionally, with less concentration of polymer, the lipid was more likely to enter the gel phase and stop phase separation from continuing to occur at low temperatures.

In the plot \ref{fig:temp_conc} the total area of these hybrid membranes (composed of PBD$_5$-$b$-PEO$_5$ is shown over the heating and cooling process. At 10 mol \%, there is strong hysteresis in the area between cooling and heating of the sample. While cooling, the system smoothly transitions to a gel phase. However, when heating, the gel phase does not melt until above 2.5 $k_BT$.  The gel phase dramatically slows the dynamics of lipids, thus, the lack of macroscopic phase separation at low polymer concentration  can be partially attributed to kinetic trapping. 

\begin{figure}
    \centering
    \includegraphics[width=0.5\linewidth]{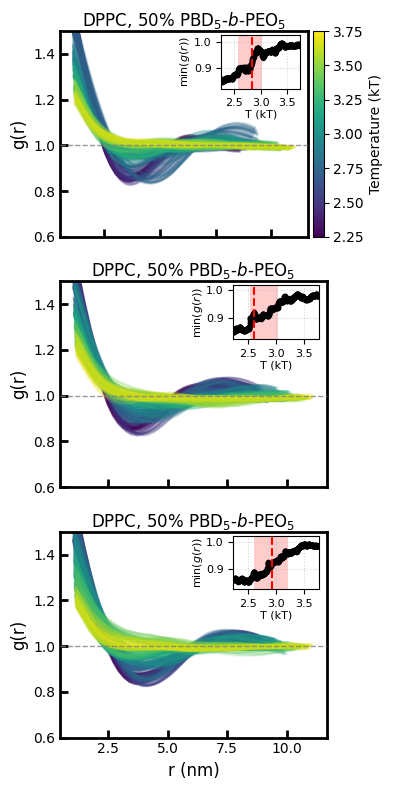}
    \caption{Pair correlation functions between polymer species of three 50 mol\% polymer simulations randomly initialized and performed in replicate. Green lines indicate values taken from high temperature, and purple lines indicate correlation functions from the lowest temperatures. Inset is a plot of the minimum value of the pair correlation function.}
    \label{fig:rdf}
\end{figure}

One feature which highlights that phase separation is occurring is the lipid-polymer pair correlation function. In Figure \ref{fig:temp_conc} and Figure \ref{fig:rdf}, it can be seen that there are long-range minima and maxima at low temperature, indicating a region of local enrichment followed by a region of depletion and then another region of enrichment, strongly indicative of phase separated domains. This demonstrates that other domains are present and a preferred length scale (due to either physical features or the finite size of the simulation box) is emerging. At high temperatures, however, there are no long range features and only a moderate increase in polymer-polymer correlations at short length scales, which is partially caused by polymer connectivity.

\FloatBarrier
\subsubsection{Interfacial Simulation with DPPC}
\begin{figure}
    \centering
    \includegraphics[width=1\linewidth]{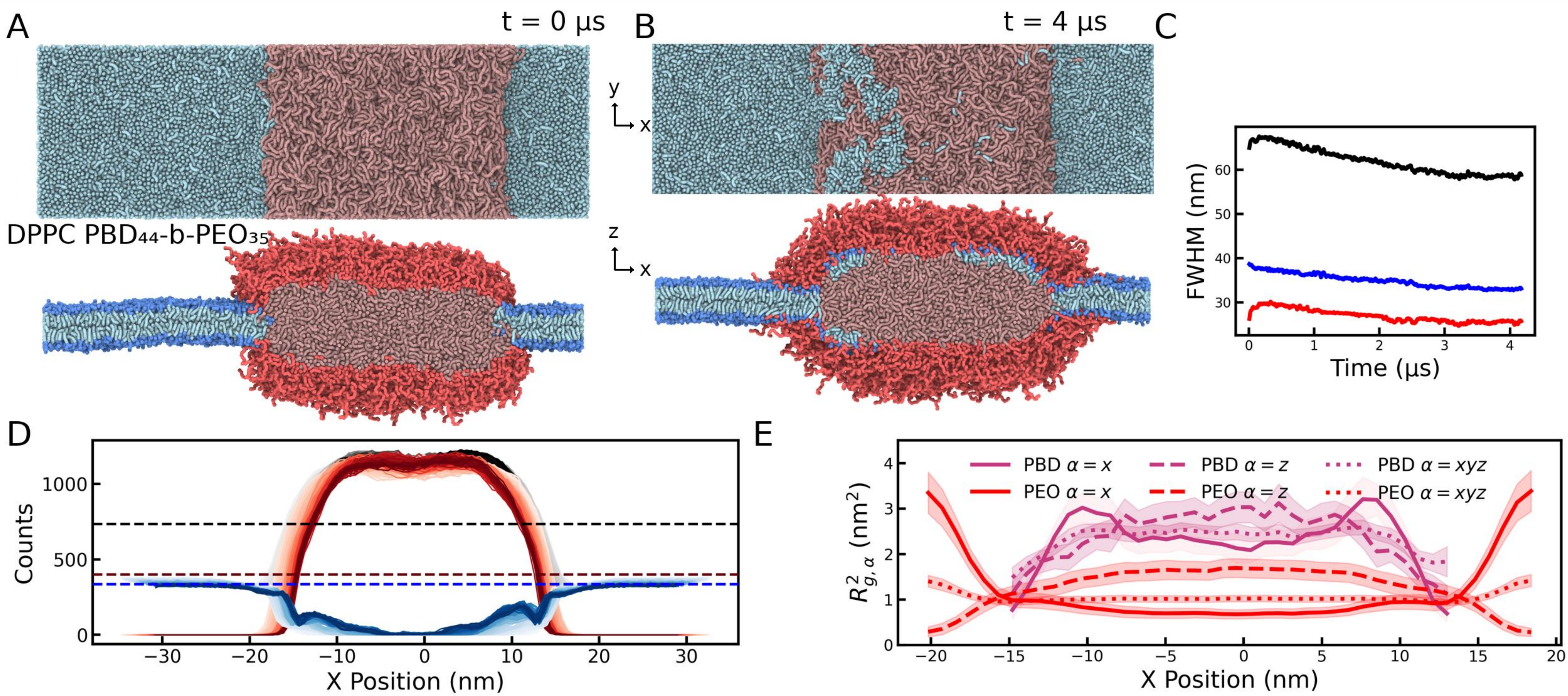}
    \caption{DPPC and PBD$_{44}$-b-PEO$_{35}$ hybrid membrane with 30 mol\% polymer simulated at 2.47 kT for 4 $\mu s$ under NP$_{xz}$LT ensemble. (A) Morphology of the system after initial equilibration, where the top snapshot shows a top down view of the membrane not showing the hydrophilic beads, and the second snapshot shows a side view of the membrane. (B) Morphology of the system after 10 $\mu$s of simulation from top down and side view. (C) Full width half max of polymer domain and lipid domain as a function of time (D) Concentration of lipids (red), polymers (blue), and the sum of lipids and polymers (black) over the last 2 $\mu$s as a function of the x coordinate, histograms are generated over a window of five simulation frames. Dashed lines shows the concentration of a pure lipid bilayer (blue), pure polymer membrane of the same PBD and PEO block lengths (red) and sum of pure lipid and pure polymer bilayers (black). (E) Components of the radius of gyration tensor of all polymers, in the x (solid line) and z (dashed line) dimension and total (dotted line), binned as a function of the x coordinate, for PEO (red) and PBD (purple) separately, averaged over the last 2 $\mu $s.}
    \label{fig:interface_dppc}
\end{figure}

In the main manuscript, we showed and discussed the phase separation of a DOPC and PBD$_{35}$-$b$-PEO$_{24}$. We also performed simulations of a DPPC and PBD$_{44}$-b-PEO$_{35}$ membrane, showin in Figure \ref{fig:interface_dppc}. First, it is important to realize that this simulation, unlike the one in the main manuscript, is not unquestionably equilibrated. The principal quantities used to understand the degree of equilibration were the full width half max of the lipid and polymer domains, as well as the   sum of the polymer and lipid FWHM. The DOPC membrane had two advantages. First, a shorter PEO chain was used which meant less water needed to be simulated to prevent the simulation from seeing itself over finite boundary conditions. Second, DOPC has significantly faster kinetics than DPPC because it is a more fluid lipid. Nevertheless, there is still interesting behavior from this snapshot in both how it is similar and different to the case of DOPC.

First, the concentration of the lipids is much lower on the membrane surface. This could be due to equilibration or due to a increased partitioning of lipids to the pure-lipid phase and away from the membrane surface. DOPC has less order and less hydrophobicity difference with PBD compared to DOPC, both suggest that the concentration of lipids atop the polymer phase should be higher in DOPC than DPPC. 
Second, the interface is still unzipped or partially unzipped. Because of the proximity of this interface to the pure lipid phase, this behavior is always to be expected. 

Finally, there is not as prominent of an increase in thickness near the interface. Examining the polymer chains, it can be observed that the PBD is stretched in the x dimension but compressed in the $z$ dimension near the interface. This effect might be caused by the increased bending stiffness of DPPC lipids and prefer straighter interfaces, as well as the reduced equilibration time.

\subsection{DPPC and PE-$b$-PEO}
In the Martini parameterization, the tails of lipids are parameterized with $C1$ beads, which are the most hydrophobic bead type. Chemically, lipid tails are saturated carbon chains, technically equivalent to polyethylene chains. As such, we create a polymer with no hydrophobicity mismatch which is chemically similar to polyethylene, where each bead represents two monomers.

\begin{figure}
    \centering
    \includegraphics[width=0.7\linewidth]{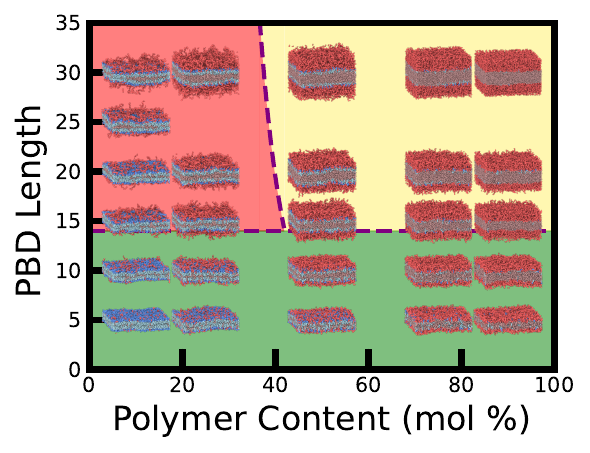}
    \caption{Morphologies of hybrid DPPC (L$_\alpha$)--block co-polymer membranes as a function of polymer content and PE number of beads (note that each bead represents 2 PE monomers). Shading of squares is predictions from theoretical expectations, where the boundaries between are determined based on a PEO length of 10 monomers. Green regions: mixed morphology, blue: lateral phase separation, red: unzipping, and yellow: polymer rich morphologies. Snapshots show morphologies observed from simulations at given state points. Light blue: lipid tails, Dark blue: lipid head groups, Light red: PE, dark red: PEO.}
    \label{fig:pepeo_snapshots}
\end{figure}

The morphologies of simulations in this system elucidate the physics of the system, as shown in Figure \ref{fig:pepeo_snapshots}. Namely, at high polymer content and long polymers, we see either mixing or polymer rich membranes. Lipids never exist in the interior of a thick membrane. Lateral phase separation is never observed. Finally, if there is not enough polymer to span the membrane, behavior similar to unzipping (or vertical phase separation) is still observed. The features are less prominent without hydrophobicity mismatch. These observations suggest that the different head groups and molecular architectures can create several morphologies without hydrophobicity mismatch.
\subsection{DPPC and 1,2 PBD-$b$-PEO}
Two previously published models for 1,2 PBD-$b$-PEO exist, the first was proposed by Vreeker et al.\cite{vreeker2025nanopore,schlitter2025nanoscale} (Model 1) and the second by Muller et al. \cite{Muller2023} (Model 2). In the main text, we discuss model 1, as it has similar behavior to 1,4 PBD-$b$-PEO. As shown below, model 2 only displays mixing behavior because the polymer beads have reduced self interactions. As such, we include the simulations of Model 2 in the SI.

\FloatBarrier
\subsubsection{Model 1}

\begin{figure}
    \centering
    \includegraphics[width=0.5\linewidth]{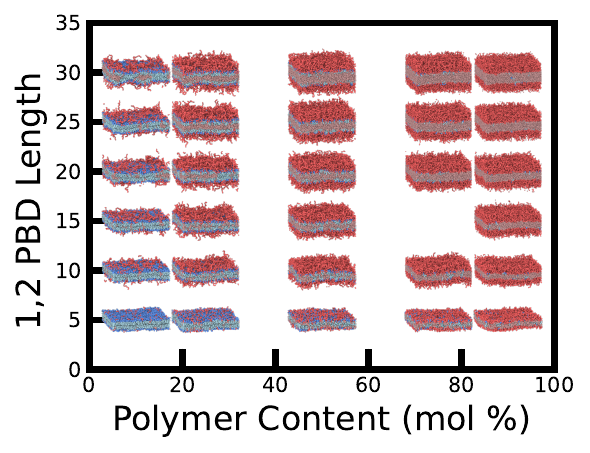}
    \caption{Morphologies of hybrid DPPC (L$_\alpha$)--block co-polymer membranes as a function of polymer content and 1,2 PBD length \cite{vreeker2025nanopore}. Snapshots are shown atop their respective conditions, showing morphologies observed from simulations at given state points. Light blue: lipid tails, Dark blue: lipid head groups, Light red: PE, dark red: PEO.}
    \label{fig:vreeker-snapshots}
\end{figure}

In the main text, we discuss how with model 1, mixing, lateral phase separation, and polymer-rich membranes can be observed. In Figure \ref{fig:vreeker-snapshots}, we show the morphologies for all PBD-$b$-PEO + DPPC systems. Because we did not develop a model for the thickness of this polymer as a function of the PBD and PEO blocks, we did not draw the lines on the phase diagram. Nevertheless, again, in the top left, unzipped morphologies are seen. In the top right, polymer rich morphologies are seen, and at the bottom, mixing is seen. With short polymers and about 50 mol \% lipid, some correlations appear between lipids and polymers, however, it doe not represent the strong phase separation seen in the 1,4 PBD system. The origin of this difference is the reduced hydrophobicity of the polymer as well as the differently sized beads that can pack in unique configurations.

\FloatBarrier
\subsubsection{Model 2}

\begin{figure}
    \centering
    \includegraphics[width=0.5\linewidth]{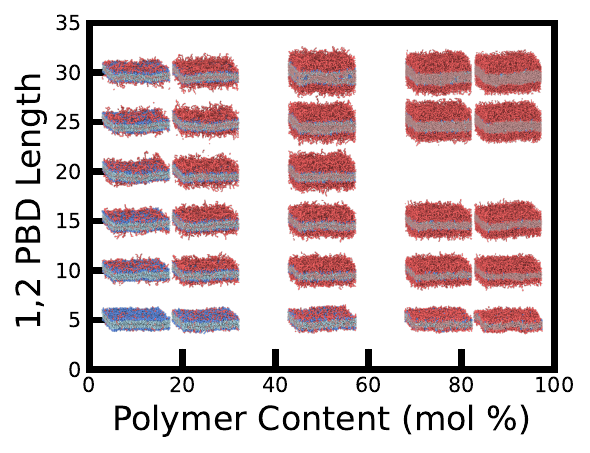}
    \caption{Morphologies of hybrid DPPC (L$_\alpha$)--block co-polymer membranes as a function of polymer content and 1,2 PBD length \cite{Muller2023}. Snapshots are shown atop their respective conditions, showing morphologies observed from simulations at given state points. Light blue: lipid tails, Dark blue: lipid head groups, Light red: PE, dark red: PEO.}
    \label{fig:muller-snapshots}
\end{figure}

Model 2 only shows mixing and polymer-rich morphologies (Figure \ref{fig:muller-snapshots}). Inside some polymer-rich morphologies, micelles form because the reduced self-interaction of the polymer results in an effective attraction between lipids and polymers. With a dilute concentration of very long polymer, the membrane still has the thickness of the lipid membrane and there is no variance in the thickness across the membrane. These unique behaviors arise due to the reduced self interactions which is a nonphysical interaction for the systems we are modeling.

\begin{figure}
    \centering
    \includegraphics[width=0.5\linewidth]{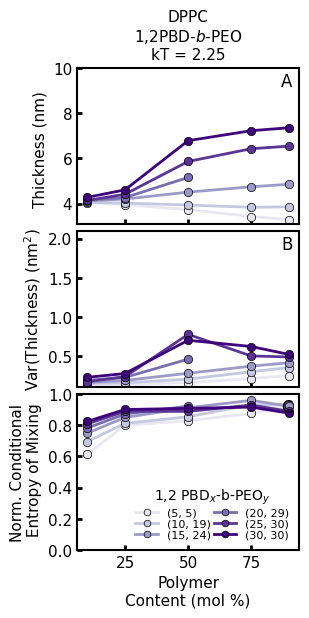}
    \caption{Properties of hybrid membranes for DPPC and 1,2 PBD-$b$-PEO \cite{Muller2023}, as function of polymer content (x axis) and hydrophobic chain length (color). (A) Thickness of the hydrophobic core of the membrane. (B) Spatial variance of thickness of hydrophobic core of the membrane. (C) Normalized conditional entropy of mixing between hydrophobic lipid tail beads and hydrophobic polymer beads.}
    \label{fig:muller-cond-ent}
\end{figure}

Additionally, we show the same properties we analyzed in the main text for this 1,2 DPPC system. The behavior is consistent with only seeing mixing and polymer rich membranes. Namely, the conditional entropy of mixing is very high. There is a peak in the variance of thickness, however, this does not correspond to the previously discussed unzipping behavior. This peak originates from lipids forming micelles inside of the polymer membrane because of the strong cross-interaction between the lipid and polymer phases.

\FloatBarrier
\subsection{DOPC and PBD-$b$-PEO}

Finally, we examine a system with 1,4 PBD-$b$-PEO and DOPC. The parameters of the Flory Huggins spinodal were obtained by using the average attractive well depth for 3 C1 type beads and 1 C4h type beads. As such, the predicted region of demixing is smaller than the case with DPPC. The other lines are adopted from the thickness of the DOPC membrane which is thinner than a DPPC membrane, stabilizing a larger region of unzipping behavior. 

\begin{figure}
    \centering
    \includegraphics[width=0.5\linewidth]{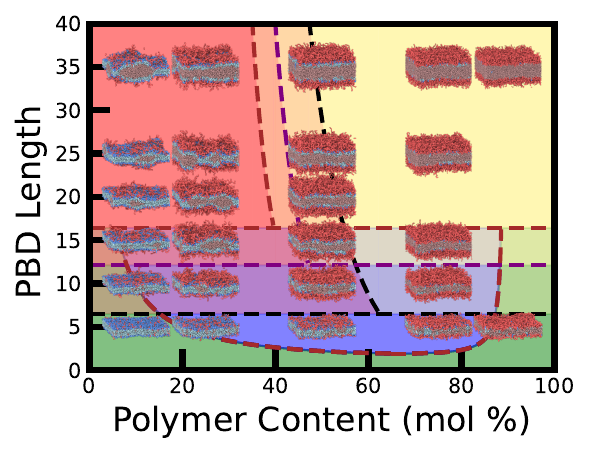}
    \caption{Morphologies of hybrid DOPC (L$_\alpha$)--block co-polymer membranes as a function of polymer content and 1,4 PBD length \cite{Muller2023}.Shading of squares is predictions from theoretical expectations, where the boundaries between are determined based on a PEO length of 10 monomers. Green regions: mixed morphology, blue: lateral phase separation, red: unzipping, and yellow: polymer rich morphologies.  Snapshots are shown atop their respective conditions, showing morphologies observed from simulations at given state points. Light blue: lipid tails, Dark blue: lipid head groups, Light red: PE, dark red: PEO.}
    \label{fig:dopc_snapshots}
\end{figure}

In Figure \ref{fig:dopc_snapshots}, the morphologies of all performed simulations are shown. In the top left, large unzipped domains are seen. These domains have more prominent globules of lipid than DPPC, likely caused by the reduced monolayer bending stiffness for DOPC as compared to DPPC (and seen in the interfacial simulations. The transition from unzipping to polymer-rich morphologies is predicted accurately by the thickness characteristic. 

Similar to the DPPC phase diagram, at the triple point (between lateral phase separation, polymer-rich and unzipping), the morphologies are hard to distinguish as there is character of all three. With short polymers, there is lateral phase separation, however, the concentration of DOPC within the polymer-phase is much higher than in the DPPC phase diagram.
\FloatBarrier
\bibliography{bib}